\documentclass[12pt,onecolumn,journal]{IEEEtran}
\usepackage[margin=2.54cm, lmargin=2.54cm]{geometry}
\usepackage{float}
\usepackage{amsfonts}
\usepackage{amsbsy}
\usepackage{amssymb}
\usepackage{times}
\usepackage{graphicx}
\usepackage{enumerate}
\usepackage[usenames]{color}
\usepackage[dvips]{pstcol}
\usepackage{epstopdf}
\usepackage{cite}
\usepackage{amssymb}
\usepackage{amsfonts}
\usepackage{graphicx}
\usepackage{epsfig}
\usepackage{psfrag}
\usepackage{xcolor}
\usepackage{amsfonts, bm}
\usepackage{epstopdf}
\usepackage{cite}
\usepackage{color}
\usepackage{xcolor}
\usepackage{subfig}
\usepackage{verbatim}
\usepackage{multirow}
\usepackage{array}
\usepackage{booktabs}
\usepackage{amsthm}
\usepackage{makecell}

\usepackage[linesnumbered, ruled]{algorithm2e}
\usepackage{algpseudocode}
\usepackage{amsmath}

\newcommand{\bH}{\mathbf{H}}

\newcommand{\bF}{\mathbf{F}}

\newcommand{\bV}{\mathbf{V}}

\IEEEoverridecommandlockouts

\begin{document}
	
	\title{SCAN: Semantic Communication with Adaptive Channel Feedback}
	\author{\IEEEauthorblockN{ Guangyi Zhang, Qiyu Hu, Yunlong Cai, and Guanding Yu   }
		\thanks{ G. Zhang, Q. Hu, Y. Cai, and G. Yu are with the College of Information Science and Electronic Engineering, Zhejiang University, Hangzhou 310027, China (e-mail: zhangguangyi@zju.edu.cn; qiyhu@zju.edu.cn; ylcai@zju.edu.cn; yuguanding@zju.edu.cn).  }  }
		\maketitle
	\vspace{-3.3em}
	\begin{abstract}
	In existing semantic communication systems for image transmission, some images are generally reconstructed with considerably low quality. As a result, the reliable transmission of each image cannot be guaranteed, bringing significant uncertainty to semantic communication systems. To address this issue, we propose a novel performance metric to characterize the reliability of semantic communication systems termed semantic distortion outage probability (SDOP), which is defined as the probability of the instantaneous distortion larger than a given target threshold. Then, since the images with lower reconstruction quality are generally less robust and need to be allocated with more communication resources, we propose a novel framework of Semantic Communication with Adaptive chaNnel feedback (SCAN). It can reduce SDOP by adaptively adjusting the overhead of channel feedback for images with different reconstruction qualities, thereby enhancing transmission reliability. 
	To realize SCAN, we first develop a deep learning-enabled semantic communication system for multiple-input multiple-output (MIMO) channels (DeepSC-MIMO) by leveraging the channel state information (CSI) and noise variance in the model design. 
	We then develop a performance evaluator to predict the reconstruction quality of each image at the transmitter by distilling knowledge from DeepSC-MIMO. 
	In this way, images with lower predicted reconstruction quality will be allocated with a longer CSI codeword to guarantee the reconstruction quality.
	We perform extensive experiments to demonstrate that the proposed scheme can significantly improve the reliability of image transmission while greatly reducing the feedback overhead.

	\end{abstract}
	%\vspace{-1.3em}
	

	\IEEEpeerreviewmaketitle
	
	\section{Introduction}
	Driven by the extensive deployment of various intelligent services, such as the autonomous driving and the Internet of Everything \cite{gunduz_survey, NiuKai,Yang}, a fierce demand for transmitting massive amounts of information has sprung up.     
	It pushes industry and academia to further improve the transmission efficiency. 
	To this end, semantic communication, as a new communication paradigm, has been regarded as a promising technology in 6G communications. 
	Semantic communication is not the pursuit of accurate bit transmission, but the ability to transmit the desired meaning of the message, resulting in higher transmission efficiency than traditional communication system based on Shannon theory \cite{Yang}. 
	%It can significantly reduce the resource requirements of wireless communication systems, such as power and bandwidth. 
	Therefore, the research on the semantic-aware physical layer communication design for semantic communication is expected to be explored.

	\subsection{Prior Work}
	Recently, inspired by the success of deep learning, autoencoder architectures parameterized by deep neural networks (DNNs) have been used to implement semantic communication systems,  achieving  significant performance gains \cite{Yang}. %{JSCC, UDeepSC,JSCCf, vqvae,ADJSCC,PC,CurrLearning, execessSeman,zhaohui,OFDMJSCC,Constellation,Revise,T1,T3}yifei yuan. 
	In particular, the existing focuses of semantic communication can be mainly divided into two categories: (i) Design effective semantic encoding and decoding algorithms \cite{JSCC,Speech, UDeepSC,JSCCf, vqvae,ADJSCC,PC,CurrLearning,T2, execessSeman}; (ii) Investigate advanced physical layer modules for semantic communication \cite{Revise,zhaohui,OFDMJSCC,Constellation,T1,T3}.
	For the first issue, the transmitter and receiver are regarded as a pair of encoder and decoder consisting of DNNs, which are exploited to directly encode the input data of different modalities to channel symbols at the transmitter and decode the received channel symbols at the receiver. 
	The deep joint source and channel coding (JSCC) technique for wireless image transmission has been firstly proposed in \cite{JSCC}, where the image pixel values are mapped to the complex-valued channel symbols through a well-designed encoder. An attention-based semantic communication system for speech transmission has been proposed in \cite{Speech}. For multi-modal data transmission, a unified joint source-channel coding semantic communication system for multi-modal data has been proposed in \cite{UDeepSC}. 
	For the second issue, the researchers dedicated to implementing semantic communication by revising or redesigning the modules in conventional communications \cite{Revise}. It is more adaptive to channel variations by considering both the semantic information of source data and channel state information (CSI).
	In \cite{zhaohui}, the problems of resource allocation and semantic information extraction for wireless semantic communication with rate splitting have been investigated. In \cite{OFDMJSCC}, orthogonal frequency division multiplexing (OFDM) has been combined with an autoencoder for wireless image transmission over multipath fading channels, where the multipath channel and OFDM are represented by differentiable layers so that the system can be trained in an end-to-end manner. Moreover, a semantic-driven constellation design has been considered in \cite{Constellation} to improve the reconstruction quality of JSCC.
	%In general, semantic communication can be implemented by redesigning or revising the modules in the physical layer communications to achieve higher performance \cite{Revise}.
	
	Although aforementioned semantic works have achieved significant performance gains, most of them only consider the single-input single-output (SISO) channels. However, when implementing the semantic communication systems to the multiple-input multiple-output (MIMO) scenarios, there will be a number of new issues to solve. Among them, channel feedback for semantic communication with MIMO is of great importance. Specifically, the strengths of MIMO, e.g., high spectral efficiency, are highly dependent on the acquisition of CSI by the base station (BS), and require user equipment to feed the CSI back to the BS through feedback links, especially in frequency division duplex (FDD) scenarios  \cite{CSISurvey,AdaptMIMO2,ModelPHY}. 
	Nevertheless, the substantial antennas at the BS for massive MIMO lead to a huge dimensionality of the CSI matrix, which dramatically increases the feedback overhead. To address this issue, many techniques have been developed to reduce the overhead of channel feedback, such as vector quantization and codebook-based approaches \cite{codebookCSI}. Based on compressive sensing (CS), several algorithms have been developed to compress the CSI matrix to reduce the overhead \cite{AMP,CSCSI}, which outperform the quantization-based methods  by using the spatial and temporal correlation of CSI. 
	In particular, by transforming the CSI matrix into a sparse domain, low-dimensional compressed codewords can be obtained for feedback.
	The authors in \cite{CSCSI} employed the spatial correlation between proximate antennas to compress the CSI matrix in the sparse spatial-frequency domain. In addition, there have also been many deep learning-based methods \cite{csi_cnn,  CSIPaper1, CLNET,multirate, T4} that use DNNs to compress the CSI matrix, such as CsiNet \cite{CSIPaper1} and CLNet \cite{CLNET}. 
	These methods adopt an autoencoder structure consisting of an encoder and a decoder, which outperform the CS-based methods with a much-reduced computational complexity. 
	Specifically, the authors in \cite{CLNET} proposed a forged complex valued input layer to process signals and utilized spatial-attention to enhance the performance of the network. For variable-rate feedback, a multi-rate framework has been developed in \cite{multirate} to
	compress the CSI matrix with different compression ratios.
	However, despite the satisfactory performance, these methods may not be optimal in semantic communications since they have not considered semantics of sources in the designs.
	
	\subsection{Motivation and Contributions}
\begin{figure*}[t]
		\begin{centering}
			\includegraphics[width=0.5\textwidth]{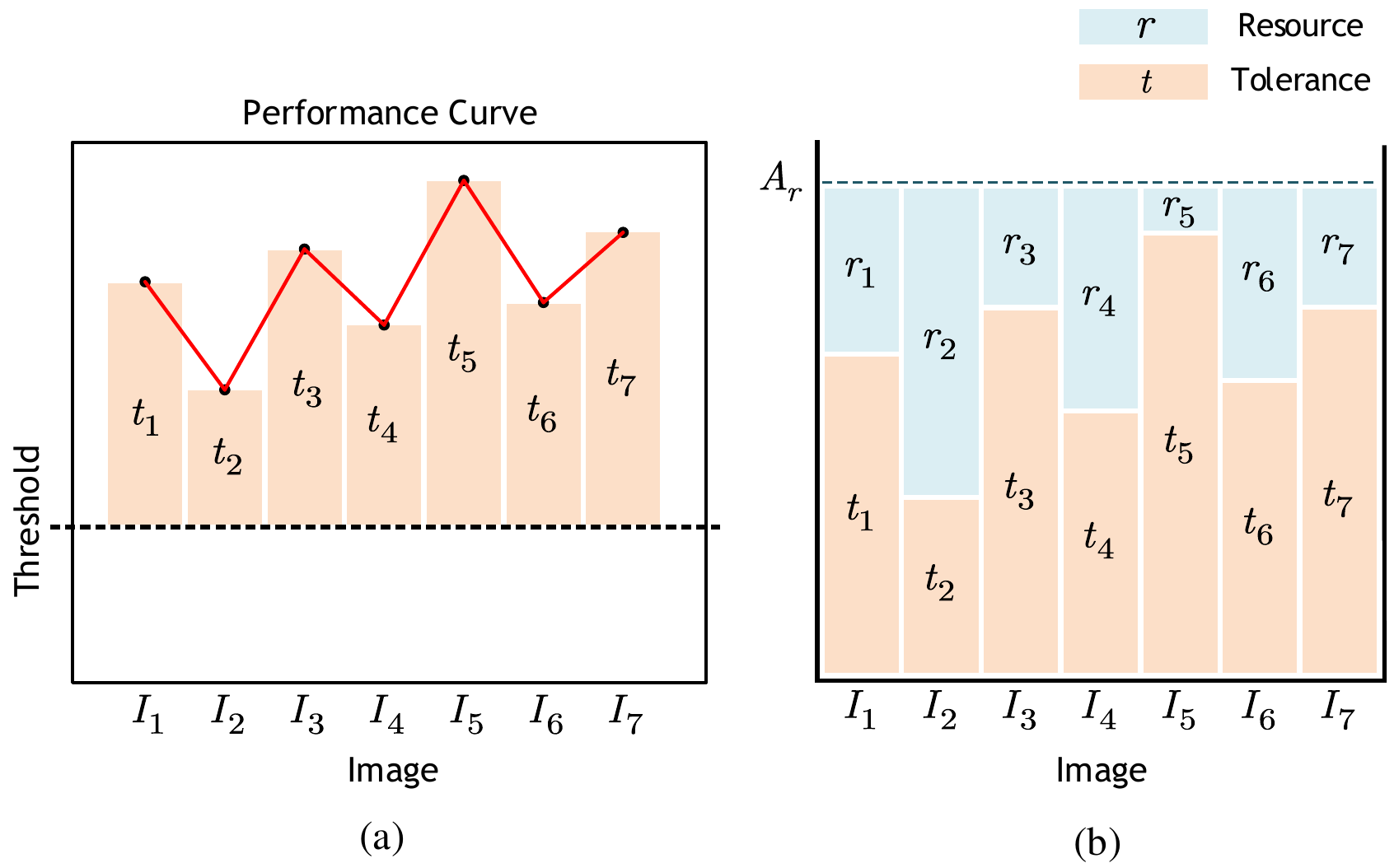}
			\par\end{centering}
		\caption{(a) The target PSNR threshold and tolerance; (b) Illustration of the water-filling.}
		\label{ProImage}
	\end{figure*}

	To the best of our knowledge, semantic communication systems for image transmission generally achieve quite different performance on different images even in error-free transmission, as shown in Fig. \ref{ProImage}(a). This is mainly because the images are of different complexities and the capabilities of the DNN models to handle different images are generally different. 
	Moreover, we observe that if the images are required to be reconstructed with a reconstruction quality larger than a given threshold, the images with lower reconstruction quality are generally less tolerant to disturbances. In this case, when the channel condition is poor, some images will be reconstructed with rather low quality, leading to high uncertainty and lack of performance guarantee. 
	Hence, it is necessary to allocate proper resources, such as channel feedback bits and transmission power, to the images with low reconstruction quality to guarantee their reconstruction qualities thereby enhancing the reliability, as shown in Fig. \ref{ProImage}(b).
	However, existing semantic communication systems only focus on optimizing the end-to-end average distortions, such as average peak-signal-to-noise (PSNR), and have not considered the reconstruction quality of each image as well as the transmission reliability.
	%Besides, noting that existing semantic communication systems only consider the SISO channel with additive white Gaussian noise (AWGN), it is necessary to extend the semantic communication system to the MIMO scenarios with the consideration of physical layer design, such as channel feedback and precoding. 
	Therefore, in this paper, we investigate the adaptive channel feedback design of image semantic communication systems with MIMO to improve the transmission reliability.

	Unlike previous works that consider the end-to-end expected distortion, we first propose a novel metric, semantic distortion outage probability (SDOP), defined as the probability of the distortion greater than a target distortion threshold. It is employed to characterize the reliability of semantic communication for image transmission. Based on the observation that images with low reconstruction quality are less robust when targeted to exceed the PSNR threshold\footnote{PSNR can be viewed as the inverse of distortion, hence ``targeted to exceed the PSNR threshold" actually denotes ``targeted to not exceed the distortion threshold”. In the following, we will use both statements for clarity.}, we develop a novel framework of Semantic Communication with Adaptive chaNnel feedback (SCAN) for image transmission to reduce SDOP thereby enhancing transmission reliability. To realize SCAN, we first propose an attention mechanism-based MIMO transmission semantic communication system (DeepSC-MIMO), which can adapt to channel and noise variations. In particular, the CSI matrix and noise variance are adopted to generate the attention mask to adaptively allocate higher power to more important feature vectors. 
	
	In order to know the reconstruction quality in advance, we propose a performance evaluator at the transmitter. It takes the image, CSI matrix, and noise variance as input and outputs the predicted reconstruction quality. To train the evaluator, a knowledge distillation loss is proposed by increasing the similarity of output features between the performance evaluator and DeepSC-MIMO. Then, an instance-wise adaptive scheme can be developed to adjust the length of the CSI codeword for each image, where the transmitter determines the compression level of the CSI matrix based on the predicted reconstruction quality of a given image. The images with lower predicted reconstruction quality will be allocated with a longer CSI codeword to guarantee the reconstruction quality, while the codeword length of those with higher predicted reconstruction quality can be reduced instead. Furthermore, a group-wise adaptive algorithm is developed to simultaneously determine the CSI compression levels for a group of images to meet the requirement of a group of images waiting for transmission.   
	Our simulation results show that the proposed SCAN can significantly improve the SDOP performance while greatly reducing the feedback overhead.
	The main contributions of this paper are summarized as follows.
	\begin{itemize}
		\item A novel performance metric, SDOP, is proposed to characterize the reliability of an end-to-end semantic communication system. 
		
		\item  We propose DeepSC-MIMO, which adapts to channel and noise variations by leveraging the CSI and noise variance based on the attention mechanism.

		\item A performance evaluator is developed based on knowledge distillation \cite{hint}, where a novel distillation loss is proposed to distill the knowledge contained in the semantic model.
		
		\item Based on the predicted performance, an adaptive instance-wise channel feedback scheme is proposed to adjust the compression level of the CSI matrix. 
		
		\item We further propose a group-wise adaptive algorithm, which aims to determine the CSI compression levels of a group of images to optimize the SDOP.
		
	\end{itemize}

	\subsection{Organization and Notations}
	The rest of this paper is structured as follows. Section \ref{Definition} introduces the definition of SDOP. The framework of the proposed SCAN is presented in Section \ref{SCANFramework}. In Section \ref{PEdesign}, the knowledge distillation-based performance evaluator is described. Followed by the corresponding module designs of SCAN, the instance-wise and group-wise adaptive designs are given in Section \ref{DetailDesign}. Simulation results are presented in Section \ref{Simulation}. Finally, Section \ref{Conclusion} concludes this paper.

	\emph{Notations:} Scalars, vectors, and matrices are respectively denoted by lower case, boldface lower case, and boldface upper case letters. For a matrix $\mathbf{A}$, ${\bf{A}}^T$, ${\bf{A}}^H$, and $\|\mathbf{A}\|_2$ are its transpose, conjugate transpose, and Frobenius norm, respectively. For a vector $\mathbf{a}$, $\|\mathbf{a}\|$ is its Euclidean norm. 
	%We use $\mathbb{E}\{ \cdot \}$ for the statistical expectation, $\textrm{Tr}\{ \cdot \}$ is the trace operation, and $| \cdot |$ denotes the absolute value of a complex scalar. 
	%and $\odot$ is the element-wise multiplication of two matrices, i.e., Hadmard product. 
	Finally, ${\mathbb{C}^{m \times n}}({\mathbb{R}^{m \times n}})$ are the space of ${m \times n}$ complex (real) matrices.

	\section{Definition of SDOP}\label{Definition}
	In this section, we propose a probabilistic model of semantic communication systems, including the definitions of mappings and variables. Then, we give the definition of the proposed SDOP based on the excess distortion event.
	\subsection{Problem Formulation}
\begin{figure}[t]
		\begin{centering}
			\includegraphics[width=0.50\textwidth]{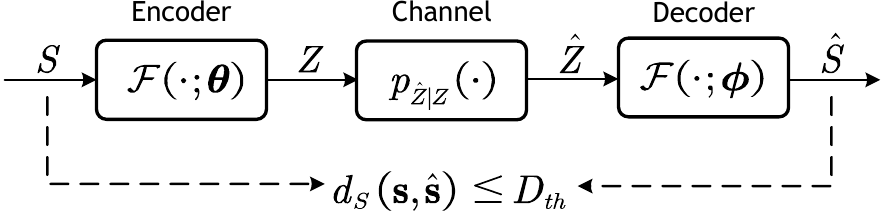}
			\par\end{centering}
		\caption{Probabilistic model of a semantic communication system.}
		\label{ProModel}
	\end{figure}

	A typical semantic communication system can be represented by the model shown in Fig. \ref{ProModel}. In particular, the semantic communication system can be viewed as an end-to-end communication system developed to incorporate the channel coding and source coding. The encoding, decoding, and transmission procedures are parameterized by the DNNs, and the system is optimized in a back-propagation manner with the data-driven method. 
	The input image is represented by a vector, $\mathbf{s} =[ s_1,s_2,...,s_N]\in \mathbb{R}^{N\times1}$, with probability distribution $p_S$, where $N$ is the length of the vector. Moreover, $\mathbf{s}$ is considered as a realization of the random variable ${S}$, in the alphabet $\mathcal{S}$.
	Denote the encoding function of the encoder as $\mathcal{F}(\boldsymbol{\cdot} \, ; {\boldsymbol{\theta}}):\mathbb{R}^{N\times1}\rightarrow \mathbb{C}^{K\times1}$, where $\boldsymbol{\theta}$ denotes the trainable parameters. The encoder directly maps $\mathbf{s}$ into the complex channel symbol vector, which is given by
	\begin{equation}
		\mathbf{z}=\mathcal{F}\left( \mathbf{s}; {\boldsymbol{\theta}}  \right) \in \mathbb{C}^{K\times1},
	\end{equation}
	where $K$ is the number of transmitted symbols and $\mathbf{z} =[ z_1,z_2,...,z_K]\in \mathbb{C}^{K\times1}$ can be viewed as a realization of the random variable $Z$, in alphabet $\mathcal{Z}$. We define the bandwidth ratios as $\rho = \frac{K}{N}$.
	Subsequently, the encoded channel symbol vector $\mathbf{z}$, is transmitted through the channel with transition probability, $p_{\hat{Z}|Z}$. Then, we obtain the received symbol vector, $\hat{\mathbf{z}} =[ \hat{z}_1,\hat{z}_2,...,\hat{z}_K]\in \mathbb{C}^{K\times1}$, which will be further processed by the decoder. Similarly, $\hat{\mathbf{z}}$ can be viewed as the realization of random variable $\hat{Z}$, in alphabet $\hat{\mathcal{Z}}$. The decoder employs the decoding function, $\mathcal{F}(\boldsymbol{\cdot} \, ; {\boldsymbol{\phi}}):\mathbb{C}^{K\times1}\rightarrow \mathbb{R}^{N\times1}$, to map $\hat{\mathbf{z}}$ into an estimate of the original signal for reconstruction, given by
	\begin{equation}
		\hat{\mathbf{s}}=\mathcal{F}\left( \mathbf{z};\boldsymbol{\phi}  \right) \in \mathbb{R}^{N\times 1},
	\end{equation}
	where $\boldsymbol{\phi}$ denotes the trainable parameters of the decoder. Additionally, $\hat{\mathbf{s}}=[\hat{s}_1,\hat{s}_1,...,\hat{s}_N]$ is regarded as a realization of the random variable $\hat{S}$, in alphabet $\hat{\mathcal{S}}$.
	
	We further define the semantic communication system as a tuple of mappings:
	\begin{equation}
		\mathcal{F}(\boldsymbol{\cdot}  ; {\boldsymbol{\theta}}): \mathcal{S} \rightarrow \mathcal{Z}, \quad \mathcal{H}(\boldsymbol{\cdot}  ;\mathbf{H},\sigma^2):\mathcal{Z} \rightarrow \hat{\mathcal{Z}}, \quad \mathcal{F}(\boldsymbol{\cdot}  ; {\boldsymbol{\phi}}): \hat{\mathcal{Z}} \rightarrow \hat{\mathcal{S}}.
	\end{equation}
	Considering the image transmission application, we further define the following block-wise quadratic-distortion measure function for the source,
	\begin{equation}
		\begin{aligned}
			& d_S: \mathcal{S} \times \hat{\mathcal{S}} \rightarrow \mathbb{R}, \quad d_{S}\left(\mathbf{s}, \hat{\mathbf{s}}\right) \triangleq \frac{1}{K} \sum_{i=1}^K d_S\left(s_i, \hat{s}_i\right). \\
		\end{aligned}
	\end{equation}

	\subsection{Outage for Semantic Communication}
	We observe that deep learning-based semantic communication systems typically achieve significantly different reconstruction quality on different input images, in which case some images will be transmitted with considerably low quality. 
	This reminds us that the average performance metrics, e.g., the average PSNR, will fail to characterize the true end-to-end performance of such semantic communication systems. The main reasons can be further summarized as follows:
	\begin{itemize}
		\item The semantic communication model for image transmission generally achieves different performance over different samples, where some images will be reconstructed with rather low quality, leading to high uncertainty and lack of performance guarantee.
		
		\item It is hard to perceive the distortion when the difference between the original image and the reconstructed image is sufficiently small. Hence, for human perception requirements, images are expected to be transmitted with distortion less than the minimum threshold.
		
		\item Since deep learning-based methods generally suffer from poor generalization ability and reliability, there is often the case that the model cannot handle images from new distributions, resulting in poor reliability. Thus, for applications with high reliability requirements, the average distortion measure may not be suitable.
	\end{itemize}

	When communicating over quasi-static fading channels at a given rate $R$, the random fading coefficients may occasionally be very small, in which case the Shannon capacity is zero. For reliable transmission over this kind of channels, it is desirable that the transmission rate is less than the channel capacity \cite{outage1,Execess1,SemanDis2}. It can be interpreted as the probability of failed transmission with high BER, referring to outage probability.
	However, the outage of conventional communication systems is not able to characterize the outage performance of semantic communication systems. It is mainly because the performance of a semantic communication system is highly related to the source content and model capability. In this case, the transmission failure event is also  source-dependent, but the conventional outage only considers channel factors. Therefore, it is necessary to define the outage in semantic communication by considering both sources and channels.

	\subsection{SDOP Definition}
	Intuitively, the event that some images cannot be reconstructed with a distortion less than the target threshold can be regarded as a transmission failure and inspires us to define the outage in semantic communications. Mirroring results of the excess distortion probability from joint source-channel coding \cite{Execess1,SemanDis2}, we propose a new metric called SDOP, which is defined as the probability that the instantaneous distortion is larger than the target quality-of-service (QoS) distortion. Revisiting the probabilistic model shown in Fig. \ref{ProModel}, the conditional probability, $p(\hat{\mathbf{s}}|\mathbf{s})$ can be expressed as $ p(\hat{\mathbf{s}}|\mathbf{s})=p_{\boldsymbol{\theta}}(\mathbf{z}|\mathbf{s} )p_{\hat{Z}|Z}(\hat{\mathbf{z}}|\mathbf{z}) p_{\boldsymbol{\phi}}(\hat{\mathbf{s}}|\mathbf{z})$, where $p_{\boldsymbol{\theta}}$ and $p_{\boldsymbol{\phi}}$ are defined by the encoder and decoder, respectively. As we adopt a deterministic DeepSC-MIMO, $p_{\boldsymbol{\theta}}$ and $p_{\boldsymbol{\phi}}$ can be viewed as a Dirac-delta function for simplicity. Thus, we can define the erroneous set of $\left(\mathbf{s}, \mathbf{z}\right)$ that violates the distortion constraints, $D_\textrm{th}$, as 
	\begin{equation}
		\mathcal{E}=\left\{\left(\mathbf{s}, \mathbf{z} \right) \in \mathcal{S}  \times \mathcal{Z}: d_S\left(\mathbf{s}, \mathcal{F}\left(\hat{\mathbf{z}}; {\boldsymbol{\phi }}\right)\right)>D_\textrm{th} \right\}.
	\end{equation}
	Therefore, the SDOP can be defined as the probability of exceeding the distortion constraint, $D_\textrm{th}$, which is expressed as 
	\begin{equation}\label{EP1}
		\mathbb{P}\{\mathcal{E}\} \! \triangleq \int_{\mathbf{s} \in \mathcal{S}} p_{S}\left(\mathbf{s}\right) \int_{\mathbf{z} \in \mathcal{E}\left(\mathbf{s}\right)} p_{\hat{Z} \mid Z}\left(\hat{\mathbf{z}} \mid \mathcal{F}\left(\mathbf{s};{\boldsymbol{\theta}}\right)\right)d\mathbf{z} d\mathbf{s},
	\end{equation}
	where $\mathcal{E}(\mathbf{s}) =\{\hat{\mathbf{z}}\in   \hat{\mathcal{Z}} :(\mathbf{s}, \hat{\mathbf{z}})   \in   \mathcal{E} \}$. 
	
	Considering a semantic communication system with MIMO, according to (\ref{EP1}), the channel transition probability $p_{\hat{Z}|Z}$, is related to the channel realization $\mathbf{H}$ and the variance of additive white Gaussian noise (AWGN)  $\sigma^2$. Therefore, denoting the distortion as a random variable, $\mathcal{D}$, it will be a random variable that depends on the channel realization, noise variance, and input image. Hence, (\ref{EP1}) can be further denoted by
	\begin{equation}\label{EP2}
		\mathbb{P}\{\mathcal{E}\}\triangleq \mathbb{P}\left\{\mathcal{D}(\mathbf{s},\mathbf{H},\sigma^2)  > D_\textrm{th}\right\},
	\end{equation}
	 representing the probability that the distortion is larger than the threshold. Compared with the conventional outage, it considers both the source and channel, and thus is able to capture the true outage of the semantic communication system.

\section{Proposed Framework of SCAN} \label{SCANFramework}

	In this section, we present the framework of the proposed SCAN. The proposed SCAN consists of the DeepSC-MIMO for image transmission, the performance evaluator for performance prediction, and the channel feedback scheme for precoding.

	\subsection{DeepSC-MIMO}
	\subsubsection{Overview and Settings}
\begin{figure*}[tbp]
		\begin{centering}
			\includegraphics[width=0.98\textwidth]{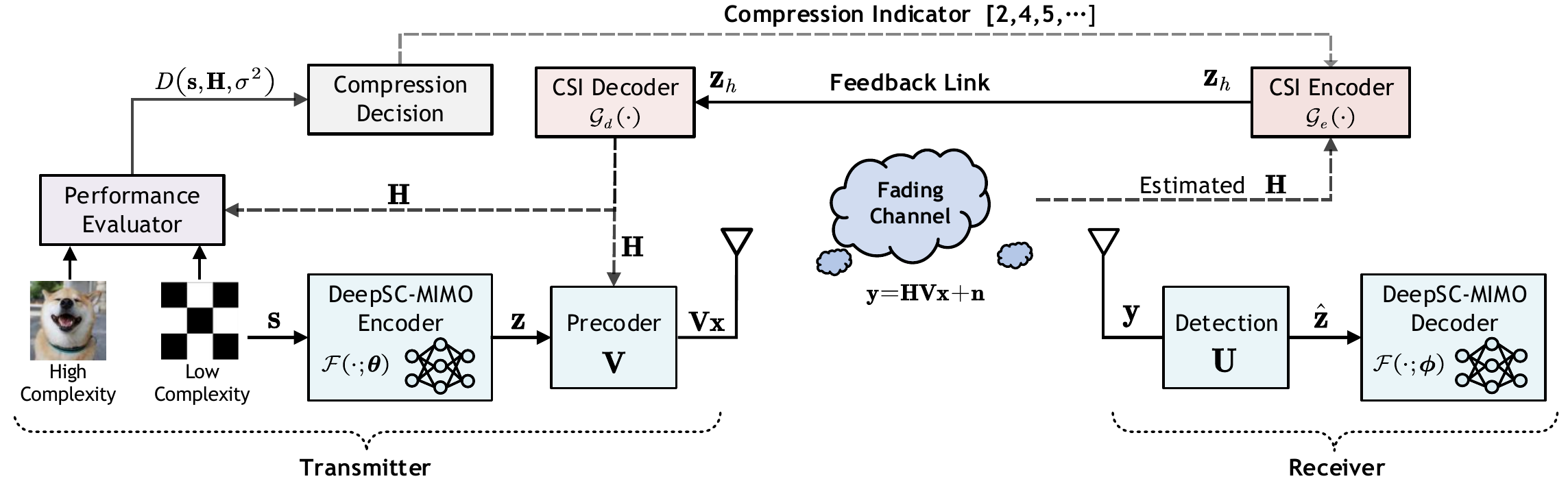}
			\par\end{centering}
		\caption{Framework of the proposed SCAN.}
		\label{SCAN_Framework}
	\end{figure*}

	%We consider a delay-constrained communication system operating over a slowly-varying fading channel. In such a scenario, it is plausible to assume that the duration of each of the transmitted codewords is smaller than the coherence time of the channel, so the random fading coefficients stay constant over the duration of each image \cite{quaisi}. 
	 By implementing the probabilistic model shown in Fig. \ref{ProModel} with actual DNNs, the DeepSC-MIMO can be carried out as shown by the DeepSC-MIMO encoder and DeepSC-MIMO decoder in Fig. \ref{SCAN_Framework}. In particular,
	 $\mathbf{s}$ is encoded directly into $\mathbf{z}$ by the encoder, which we then elaborate how to transmit to the receiver using MIMO. Let $\mathbf{V}\in \mathbb{C}^{N_r \times d}$ denote the precoder to transmit $\mathbf{z}$, where $d$ denotes the number of data streams. It can be obtained by applying the singular value decomposition (SVD) precoding with the channel matrix, $\mathbf{H}\in \mathbb{C}^{N_r \times N_t}$. Note that the image is encoded by the encoder into a $K$-dimension vector, i.e., the complex channel symbols. We split them into a number of signals, whose dimensions all equal to $d$. Taking one split signal, $\mathbf{x}\in \mathbb{C}^{d \times 1}$, as an example, we first constrain it with power, $P$, as $\|\bV \mathbf{x}\|^2 \leq P$.
 	Then, the received signal can be denoted as
	 \begin{equation}
	 	{\mathbf{y}}=\mathbf{H}\mathbf{V}\mathbf{x} + \mathbf{n},
	 \end{equation}
	 where $\mathbf{n}\in \mathbb{C}^{N_r \times 1}$ is the AWGN. At the receiver, we consider the linear receive combiner, thus the estimated signal is obtained by 
	\begin{equation}
		{\hat{\mathbf{x}}}=\mathbf{U}^{H}\mathbf{y},
	\end{equation}
	where the $\mathbf{U}$ is obtained by the receiver using $\mathbf{H}$ with SVD algorithm. Finally, the received signals are further processed by the decoder to recover the source image. 
	Moreover, the average SNR at the receiver is defined by
	\begin{equation}
		\textrm{SNR} \triangleq 10 \log _{10} \frac{\mathbb{E}\left\{\|\mathbf{H} \bV \mathbf{x}\|^2\right\}}{\mathbb{E}\left\{\|\mathbf{n}\|^2\right\}}=10\log _{10}  \frac{P}{\sigma^2}.
	\end{equation}

	\begin{figure*}[t]
		\begin{centering}
			\includegraphics[width=1.0\textwidth]{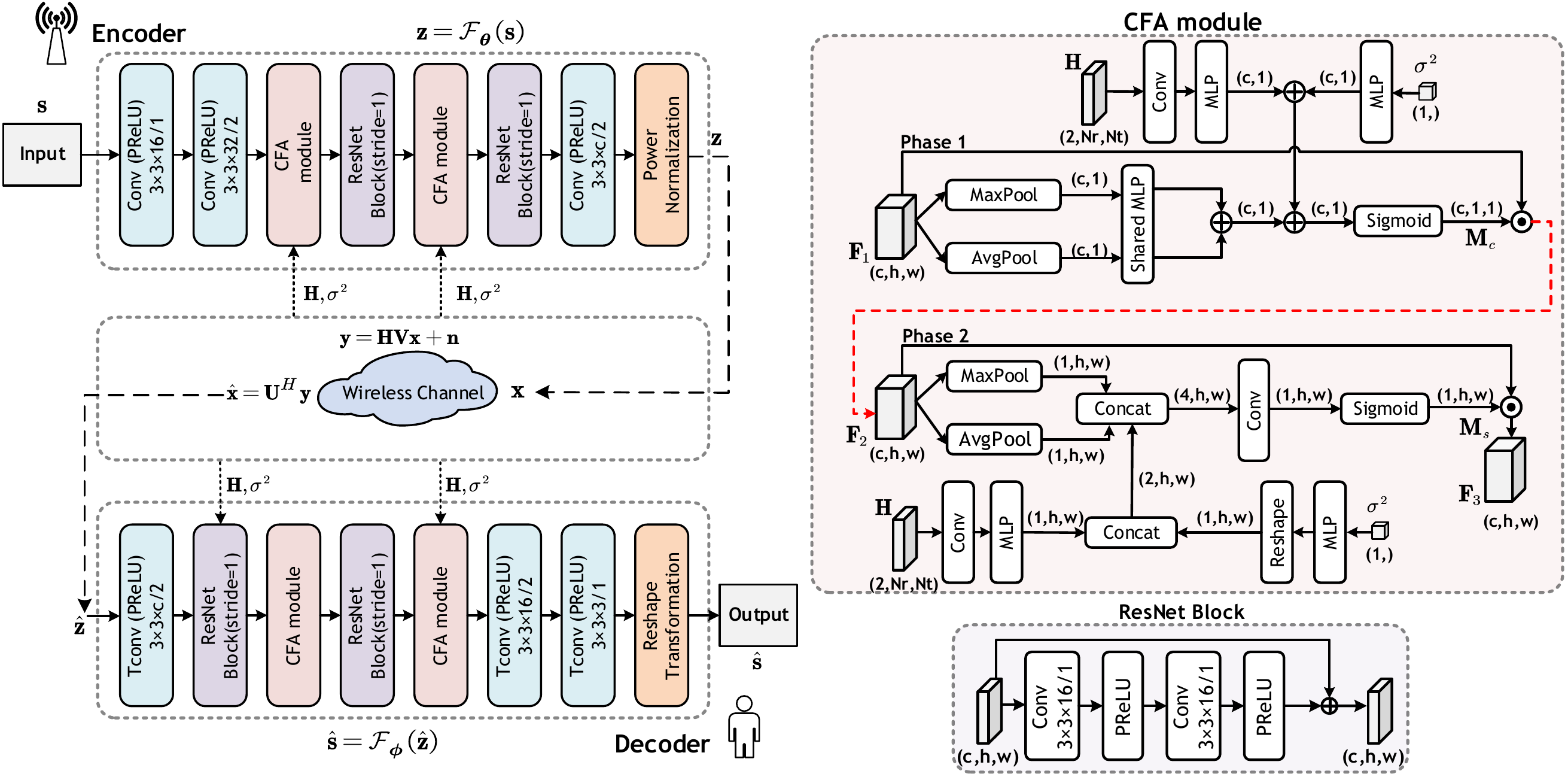}
			\par\end{centering}
		\caption{The architecture of DeepSC-MIMO with MIMO channel.}
		\label{DeMO_Model}
	\end{figure*}

	\subsubsection{Architecture Design and Training}
	The proposed DeepSC-MIMO consists of the encoder and decoder at the transmitter and receiver, respectively. We leverage the convolutional layer and residual block to design the encoder and decoder. The details are illustrated in Fig. \ref{DeMO_Model}.
	
	To make the DeepSC-MIMO adapt to different channel conditions, we develop a channel and feature attention (CFA) module. The CFA module is developed based on the channel and spatial block attention mechanism \cite{CBAM}. To make the DeepSC-MIMO adapt to the MIMO channels and improve the performance, we incorporate the CSI matrix $\mathbf{H}$ and the noise variance $\sigma^2$ into DeepSC-MIMO with a two-phase attention design, as shown in Fig. \ref{DeMO_Model}. In the first phase, $\mathbf{H}$ and $\sigma^2$ are first processed to obtain the corresponding channel features. Then, the channel features and image features are employed to generate the channel-wise attention mask, $\mathbf{M}_c \in \mathbb{R}$. Then, the mask is used to allocate different weights to different feature channels of $\mathbf{F}_1$, given by $\mathbf{F}_2=\mathbf{F}_1\odot\mathbf{M}_c$, where $\odot$ denotes Hadamard product. By integrating the CSI matrix and noise variance into the design, the DeepSC-MIMO is expected to adjust the encoder output to better fit the MIMO channel. In the second phase, we adopt $\mathbf{F}_2$, $\mathbf{H}$, and  $\sigma^2$ to generate the spatial attention mask, $\mathbf{M} $, which is used to assign different weights for the elements in each feature channel, given as $\mathbf{F}_3=\mathbf{F}_2\odot\mathbf{M}_p$. In this way, DeepSC-MIMO adapts to channel and noise variations by adaptively allocating more power to more important features according to CSI matrix and noise variance.
	
	 Assuming that each data $\mathbf{s}_{i}$ is sampled from a given dataset and CSI data $\bH_i$ is sampled from a given distribution, we input them into the encoder. Then, we obtain the encoded channel symbols, $\mathbf{z}_i=\mathcal{F}\left( \mathbf{s}_i, \bH_{i}, \sigma^2 ;{\boldsymbol{\theta}}\right)$. The transmission process is modeled as the channel layer and the received symbols can be denoted as $\hat{\mathbf{z}}_i=\mathcal{H}\left(\mathbf{z}_i ; \bH_{i},\mathbf{n}_i  \right)$, where $\mathbf{n}_i$ represents the noise sample of AWGN. Subsequently, the decoder decodes $\mathbf{z}_i$ into the reconstructed image, $\hat{\mathbf{s}}_i=\mathcal{F} \left( \mathbf{z}, \bH_{i},\sigma^2 ; {\boldsymbol{\phi}} \right)$. To jointly learn the encoder and decoder via back-propagation, we employ the mean square-error (MSE) loss, which is given by
	\begin{equation}\label{msse}
		\mathcal{L}\left(\mathbf{s},\hat{\mathbf{s}}\right)=\frac{1}{N} \sum_{j=1}^N\left(s_j-\hat{s}_j\right)^2,
	\end{equation}
	where $s_j$ and $\hat{s}_j$ denote the $i$-th elements of $\mathbf{s}$ and $\hat{\mathbf{s}}$, respectively. Moreover, the detailed training procedure is summarized in Algorithm \ref{a1}.
	
	\begin{algorithm}[t] 
		\begin{small}
			\caption{Training algorithm for DeepSC-MIMO} 
			\label{a1}
			\DontPrintSemicolon
			\SetKwInOut{Input}{Input}
			\SetKwInOut{Output}{Output}
			\renewcommand{\algorithmicrequire}{ \textbf{Input: }}     %Use Input 
			\renewcommand{\algorithmicensure}{ \textbf{Output: }}    
			\Input{The training dataset $\mathcal{S}$, consisting of the input image $\mathbf{s}_{i}$, batch size $Q$, AWGN variance $\sigma^2$, epochs $M$, and codeword length $B$.\\}
			\Output{The parameters of the trained model}
			Sample a batch of data, $\mathbf{s}_1, \mathbf{s}_2,...,\mathbf{s}_Q$.\\
			Generate a batch of CSI matrices, $\mathbf{H}_1, \mathbf{H}_2,...,\mathbf{H}_B$. from the given distribution.\\
			Generate a batch of noise samples, $\mathbf{n}_1, \mathbf{n}_2,...,\mathbf{n}_B$, according to variance, $\sigma^2$.\\
			\For{$m\leftarrow 1$ \KwTo $M$}{
				\For{$i\leftarrow 1$ \KwTo $Q$}{
					Compute encoded channel symbols $\mathbf{z}_i=\mathcal{F}\left( \mathbf{s}_i, \bH_{i},\sigma^2; {\boldsymbol{\theta}}  \right) $.\\
					Compute received symbols $\hat{\mathbf{z}}_i=\mathcal{H}\left(\mathbf{z}_i ; \bH_{i},B,\mathbf{n}_i  \right)$.\\
					Comput the reconstructed image $\hat{\mathbf{s}}_i=\mathcal{F}\left( \hat{\mathbf{z}}_i, \bH_{i},\sigma^2; {\boldsymbol{\phi}}   \right)$.\\
				}
				Calculate the average loss based on (\ref{msse}).\\
				Update the parameters of DeepSC-MIMO.
			}
		\end{small}
	\end{algorithm}

	\subsection{Design of Channel Feedback}
	%\subsection{Definition}

	\subsubsection{Encoder and Decoder for Channel Feedback}
	 Assuming perfect CSI is known by the receiver, the number of feedback elements for the considered MIMO channel should be $N_r \times N_t$ without compressing. In \cite{CSIPaper1}, it has been proven that the deep learning-based methods are more effective in dealing with the sparse input. Therefore, we sparsify the channel in the beam space domain using a $2$D discrete Fourier transform (DFT) \cite{mmWave_sparse}, which is given by
	\begin{equation}
		\tilde{\mathbf{H}}	=\mathbf{F}_{\mathrm{r}} \mathbf{H} \mathbf{F}_{\mathrm{l}}^H,
	\end{equation}
	where $\mathbf{F}_{\mathrm{r}} \in \mathbb{C}^{N_r \times N_r}$ and $\mathbf{F}_{\mathrm{l}} \in \mathbb{C}^{N_t \times N_t}$ denote the DFT matrices, respectively. Moreover, $\mathbf{F}_{\mathrm{r}}$ and $\mathbf{F}_{\mathrm{l}}$ are both unitary matrices and $\tilde{\mathbf{H}}$ is a sparse matrix. 
	Note that $\tilde{\mathbf{H}}$ is a complex matrix, we transform it into a matrix of real numbers as
	\begin{equation}
		\widetilde{\mathbf{H}}=\big[\mathfrak{Re}\big(\widetilde{\mathbf{H}}\big), \mathfrak{Im}\big(\widetilde{\mathbf{H}}\big)\big]^{T},
	\end{equation}
	where operations $\mathfrak{Re}(\cdot)$ and $\mathfrak{Im}(\cdot)$ represent obtaining the real and imaginary parts of the complex matrix, respectively. 
	
	We design the CSI encoder as $\mathcal{G}_e(\cdot)$, the CSI matrix is transformed into a $B$-dimensional codeword as $\mathbf{z}_h =\mathcal{G}_e(\widetilde{\mathbf{H}})$, where $B < 2N_r N_t$. Then, the codeword is sent to the transmitter and the CSI decoder is designed as $\mathcal{G}_d(\cdot)$, which is employed to transform the codeword into the original channel matrix, i.e., $\widetilde{\mathbf{H}} = \mathcal{G}_d(\mathbf{z}_h)$. After receiving $\widetilde{\mathbf{H}}$ at the transmitter, the CSI matrix in the spatial-angle domain can be obtained by inverse DFT. Finally, the recovered CSI matrix is used for the precoding design for DeepSC-MIMO.
	As we aim to achieve adaptive channel feedback according to the predicted reconstruction quality, the encoder needs to implement multi-rate compression. That is, the CSI encoder should be able to compress the CSI matrix into different length values.

	\subsubsection{Adaptive Channel Feedback}
	We consider the case where all transmitted images are required to be reconstructed to exceed a given minimum target PSNR threshold. It is worth emphasizing that the performance metric of image transmission usually selects PSNR, which is in fact the inversion of the quadratic-distortion. Then, the images with lower reconstruction quality, i.e., higher distortion, are generally less robust to the disturbances. This inspires us to re-determine the transmission resource allocation of the DeepSC-MIMO system by making a trade-off between the allocated transmission resource and the reconstruction quality according to the reconstruction quality of the images. 
	Based on this observation, we aim to develop an adaptive channel feedback scheme to reduce the overhead of channel feedback while maintaining a satisfactory SDOP, i.e.,  guaranteeing more images whose reconstruction quality is greater than a given target PSNR threshold.
	
	 As shown in Fig. \ref{SCAN_Framework}, the transmitter will determine the compression level of the CSI matrix based on the reconstruction quality of the given image and send a compression indicator to indicate the compression level. Then, the receiver compresses the CSI matrix according to the compression indicator, and feeds the compressed CSI codeword back to the transmitter through the feedback link for precoding. In order to know the reconstruction quality in advance so that the transmitter can determine the compression level, we propose a performance evaluator at the transmitter.  The details about the performance evaluator and the adaptive design are presented in Sections \ref{PEdesign} and \ref{DetailDesign}, respectively.

	\section{Knowledge Distillation-Based Performance Evaluator} \label{PEdesign}
	In this section, we present the detailed designs of the proposed performance evaluator based on knowledge distillation. 
	
	\subsection{Performance Influencing Factors}
	To the best of our knowledge, the reconstruction quality of each image mainly depends on three factors, image content, channel condition, and model capability. The specific explanations are listed below:
	\begin{itemize}
	\item 
	Image complexity: The knowledge about the complexity of image generally determines the image redundancy, and is of great importance in many applications. It can be used to indicate the compression ratio of an image, since images with low complexity are easier to compress than images with high complexity \cite{ImageInfo}.
	
	\item 
	Transmission errors: We consider MIMO communication, where inaccurate precoding caused by the CSI compression will negatively affect on the reconstruction quality. Nevertheless, AWGN also has an impact on the symbol transmission,  resulting in the performance degradation. Therefore, channel conditions need to be considered.
	
	\item 
	 Model capability: The information extraction ability of the model has a significant impact on the semantic communication performance of different inputs. In particular, the model tends to perform better on inputs similar to the training data. For example, if DeepSC-MIMO is trained on a series of images of dogs, there would be a performance degradation on the images of cars due to the unsatisfactory generalization ability.
\end{itemize}
	
	\begin{figure*}[tbp]
		\begin{centering}
			\includegraphics[width=0.856\textwidth]{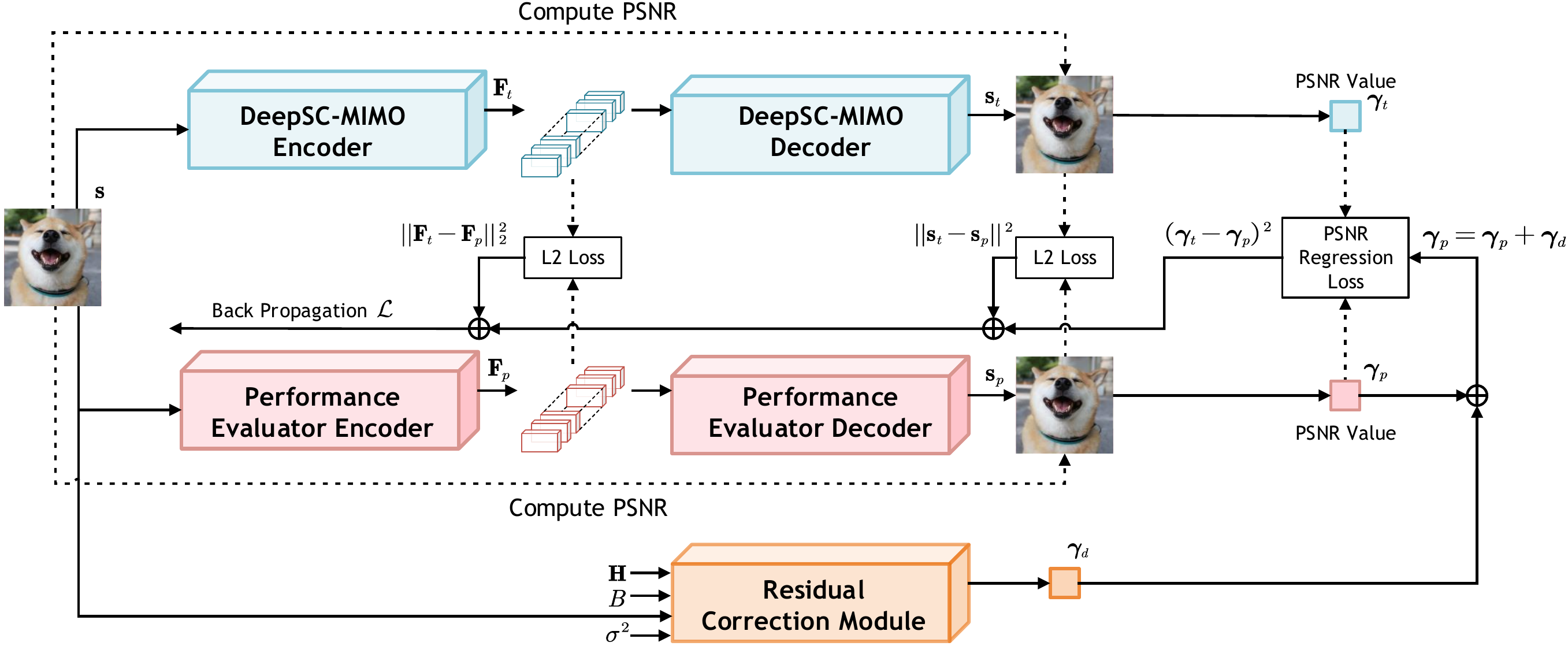}
			\par\end{centering}
		\caption{Framework of the proposed performance evaluator.}
		\label{predictor}
	\end{figure*}

	\subsection{Training with Knowledge Distillation}
	Knowledge distillation has been widely used to transfer knowledge from teacher model to student model \cite{model_distillation1,KB2,model_llation2}. Specifically, we can employ the student model to mimic the teacher model, and the output of the student model can be regarded as an approximation of the output of the teacher model.
	Thus, to predict the reconstruction quality of each given image, we propose a lightweight learnable evaluator represented by DNNs. The architecture of the performance evaluator is shown in Fig. \ref{predictor}. Similar to DeepSC-MIMO, the proposed evaluator is designed as an autoencoder structure with a residual correction module, which is beneficial to imitate the behavior of DeepSC-MIMO. 
	
	We employ the encoder and decoder parts of the performance evaluator to generate the predicted reconstructed image, $\mathbf{s}_p$, which can be regarded as an approximation of the output of DeepSC-MIMO, $\mathbf{s}_t$. Then, the reconstruction PSNR value, $\bm{\gamma}_p=\| \mathbf{s}_p - \mathbf{s}\|^2$, can be taken as the prediction of the ground-truth PSNR value achieved by DeepSC-MIMO, $\bm{\gamma}_t=\| \mathbf{s}_t - \mathbf{s}\|^2$. The training target of the proposed performance evaluator can be formulated as a regression problem, that is
	\begin{equation}\label{L}
		\min_{\bm{\pi}} \  \mathcal{L}_c \triangleq (\bm{\gamma}_t-\bm{\gamma}_p)^{2},
	\end{equation}
	where $\boldsymbol{\pi}$ denotes the trainable parameters of the performance evaluator.
	To improve the prediction accuracy,	we further develop an additional distillation loss on the intermediate features of the performance evaluator. In particular, the trained DeepSC-MIMO is set as the teacher  to help the performance evaluator to learn to better predict the reconstruction quality of the image. As presented in Fig. \ref{predictor}, the additional distillation loss is designed to encourage the features of the student model, performance evaluator, to be similar to that of the teacher model, DeepSC-MIMO. That is, we calculate the Frobenius norm of the difference between the features of the performance evaluator and DeepSC-MIMO as the measure of similarity.  Assuming that the output encoded feature of DeepSC-MIMO is $\bF_t$ and the output encoded features of the performance evaluator is $\bF_p$. Denote the output images of DeepSC-MIMO and the performance evaluator as $\mathbf{s}_t$ and $\mathbf{s}_p$, respectively. Then, the additional distillation loss is designed to increase the similarity and can be written as
	\begin{equation}\label{La}
		\mathcal{L}_p = \|\mathbf{s}_p-\mathbf{s}_t\|^2 +\|\mathbf{F}_p-\mathbf{F}_t\|_{2}^2.
	\end{equation}
	Intuitively, the additional distillation loss makes the performance evaluator learn the properties of DeepSC-MIMO, so that the performance evaluator has the similar model capability to the performance evaluator. For instance, if DeepSC-MIMO performs better on the images of dogs than on images of cats, the performance evaluator should also be so, in which way the PSNR value can be predicted more accurately.
	
	Since the PSNR value of each image is also related to the CSI codeword length and noise variance, we further introduce a residual correction module to predict the PSNR loss caused by the transmission procedure. In particular, the proposed residual correction module takes $\mathbf{s}$, $B$, and $\sigma^2$ as input, and outputs the predicted PSNR loss, which is denoted as $\bm{\gamma}_d$. Therefore, as shown in Fig. \ref{predictor}, the predicted PSNR value of the performance evaluator can be redefined as $\bm{\gamma}_p=\| \mathbf{s}_p - \mathbf{s}\|^2+\bm{\gamma}_d$. Therefore, the total loss can be denoted by
	\begin{equation}\label{Lsssa}
		\mathcal{L}=\lambda\mathcal{L}_c+\mathcal{L}_p,
	\end{equation}
	where $\lambda$ is the weighting hyperparameter.
	In order to train performance evaluator, we establish the dataset based on a trained DeepSC-MIMO. Specifically, the dataset consists of a number of tuples with the form $\left(\mathbf{s}_i,  \bH_{i}, B_i, \sigma_i^2,\bm{\gamma}^i_t, \bF_t^i,\mathbf{s}_t^i\right)$, where $\bm{\gamma}_t^i$, $\bF_t^i$, and $\mathbf{s}_t^i$ are obtained by inputting $\mathbf{s}_i$ and $\bH_i$  to DeepSC-MIMO. Moreover, with (\ref{Lsssa}), we apply the stochastic gradient descent (SGD) algorithm to update the parameters of the predictor. %The training procedure for the predictor is summarized in Algorithm \ref{a2}.
	
	\begin{comment}
	\begin{algorithm}[t] 
		\begin{small}
			\caption{Training algorithm for the performance evaluator} 
			\label{a2}
			\DontPrintSemicolon
			\SetKwInOut{Input}{Input}
			\SetKwInOut{Output}{Output}
			\SetKwInOut{Initialize}{Initialize}
			\Input{The training dataset, $\mathcal{S}$, that consists of the input images, $\mathbf{s}_{i}$, batch size, $Q$, epochs, $M$.}
			\Output{The parameters of the trained model}
			Sample a batch of data $\{ \left(\mathbf{s}_i, \bH_{i},\bm{\gamma}_t^i, \bF_p^i\right)\}_{i=1,...,B}$.\\
			\For{$m\leftarrow 1$ \KwTo $M$}{
				\For{$i\leftarrow 1$ \KwTo $Q$}{
					Compute $\bm{\gamma}_p$. \\
					Compute regression loss based on (\ref{L}).\\
					Compute distillation loss based on (\ref{La}).\\
				}
				Calculate average loss.\\
				Update the parameters of performance evaluator, $\bm{\pi}$, based on SGD.
			}
		\end{small}
	\end{algorithm}
	内容...
	\end{comment}

	\section{Proposed Adaptive Channel Feedback Design}\label{DetailDesign}
	In this section, we introduce the  proposed adaptive channel feedback design. In particular, the instance-wise adaptive design is developed to adaptively select the optimal CSI compression level based on the predicted PSNR value for each image. Moreover, the group-wise adaptive scheme is proposed to determine the CSI compression levels of a group of images within the average length constraint.
	\subsection{Instance-Wise Design}
	In the following, we focus on the limited feedback scenario, where only the compressed CSI fed back with the codeword of  a certain length is known at the transmitter. Besides, to help the transmitter predict the distortion, a very short codeword is fed back to the transmitter firstly, whose length is much less than that of really-required codewords.  In this case, the channel transformation function can be extended to $\mathcal{H}\left(\mathbf{\cdot};\mathbf{H}, B, \sigma^2 \right) $ by involving with the codeword length $B$. 
	Note that the difference between the recovered CSI matrix $\hat{\mathbf{H}}$, and real CSI matrix $\mathbf{H}$ is measured by the normalized mean squared error (NMSE) \cite{CSIPaper1}, which can be computed as 
	\begin{equation}
		\textrm{NMSE}=\mathbb{E}\left\{\frac{\|\mathbf{H}-\hat{\mathbf{H}}\|_2^2}{\|\mathbf{H}\|_2^2}\right\}.
	\end{equation}
	In particular, a smaller $B$ will lead to better feedback accuracy, i.e., higher NMSE, which will cause the performance degradation of DeepSC-MIMO. Although increasing the codeword length of the CSI matrix helps reduce the difference, it would result in high feedback overhead. Thus, there is a trade-off between the feedback overhead and system performance. 
	To realize adaptive channel feedback design, we first design an adaptive channel feedback regime with $T$-level compressed length, given as $\Lambda = \left\{ L_1, L_2,...,L_T \right\}$, where $L_1<\!L_2<\!...\!<L_T$ are the optional codeword length values. 
	
	Then, for each input image, the target can be formulated to minimize the SDOP by choosing appropriate codeword length. In fact, this can be achieved by solving the following optimization problem,
	\begin{subequations}
		\begin{eqnarray}
			\textrm{P1}: & \min & B,\\
			& \textrm{s.t.} & \hat{\mathcal{D}}(\mathbf{s},\mathbf{H}, B,\sigma^2)  \leq D_\textrm{th},  \label{c111}\\
			& & B \in \Lambda, \label{c222}
		\end{eqnarray}
	\end{subequations}
	where $\hat{\mathcal{D}}(\cdot)$ denotes the predicted distortion. In addition, constraint (\ref{c111}) ensures that the distortion of each individual reconstructed image is not greater than the target distortion $D_\textrm{th}$. Constraint (\ref{c222}) indicates that optional CSI compression levels are selected from the given set, $\Lambda$. Moreover, $\mathcal{D}(\cdot)$ is a monotonically decreasing function of $B$. In this way, according to the predicted $\hat{\mathcal{D}}(\mathbf{s},\mathbf{H}, B,\sigma^2)$, the optimal codeword length, $B^*$, can be acquired by computing the predicted performance on different codeword length values with the performance evaluator.
	In general, it is an instance-wise method, and thus the CSI codeword length for each given image is immediately available.
	
	\subsection{Group-Wise Design}
	 In addition to the instance-wise adaptive applications, there is also the case that a group of images are required to wait for transmission. Therefore, we further investigate the group-wise adaptive design aiming to ensure that more images are transmitted over a group of images with less distortion than a target threshold, i.e., minimize SDOP. Firstly, we assume that there are $M$ images, $\left\{ \mathbf{s}_m \right\}_{m=1}^{M}$, for transmission and the optional CSI codeword length values are $\Lambda = \left\{ L_1, L_2,...,L_T \right\}$, where $L_1<\!L_2<\!...\!<L_T$. We focus on minimizing the SDOP with an average codeword length constraint. Specifically, the target can be formulated as the following problem, 
	\begin{subequations}\label{P3}
		\begin{eqnarray}
			\quad \textrm{P}2: & \min& \mathbb{P}\{\mathcal{E}\} \triangleq  \frac{1}{M} \sum_{m=1}^{M}  \boldsymbol{\delta} \left(\hat{\mathcal{D}}(\mathbf{s},\mathbf{H}, B_m,\sigma^2)  > D_\textrm{th}\right) \label{f1},\\
			& \textrm{s.t.} & \frac{1}{M}\sum_{m=1}^{M} B_m  \leq L_\textrm{th}, \label{c2}\\
			& & B_m \in \Lambda,\\
			& & \mathbf{s}_m \in \mathcal{S},
		\end{eqnarray}
	\end{subequations}
	where $\boldsymbol{\delta}(\cdot)$ equals $1$ when the condition is satisfied, and $0$ otherwise. The objective function can be interpreted as the ratio of the images with a distortion above the threshold, which can be further viewed as an empirical estimation of $\mathbb{P}\left\{\mathcal{E}\right\}$. Moreover, constraint (\ref{c2}) is an average constraint ensuring that the feedback overhead is infinite, $L_\textrm{th}$ denotes the constraint satisfying $L_1 \leq L_\textrm{th} \leq L_T$.

	\begin{algorithm}[t] 
		\begin{small}
			\caption{Group-wise adaptive method.} 
			\label{A5}
			\DontPrintSemicolon
			\SetKwInOut{Input}{Input}
			\SetKwInOut{Output}{Output}
			\renewcommand{\algorithmicrequire}{ \textbf{Input: }}     %Use Input 
			\renewcommand{\algorithmicensure}{ \textbf{Output: }}    
			\Input{$M$ images, $\left\{\mathbf{s}_m \right\}_{m=1}^{M}$, optional CSI compression levels, $\Lambda = \left\{ L_1, L_2,...,L_T \right\}$, and the predicted determination of excessing the distortion event, $G_{m,l}$ for $m=1,...,M$ and $l=1,...,L$.}
			\Output{Optimized $\mathbb{P}\{ \mathcal{E} \}$ and the allocation strategy $\Omega$.}
			Initialize index list, $\mathcal{I}= \{ 1,2,...,M\}$, to indicate the index of image that has not been determined with the compression level.\\
			Initialize an empty list, $\mathcal{K}$, to indicate which image have been allocated with proper codeword length.\\
			Initialize the allocation set $\Omega[i]=  L_1 , \text{for} \ i=1,...,M$, i.e., allocates the shortest codeword length for $M$ images.  \\
			\For{$t=1,...,T$}{
				\For{$i$ in $\mathcal{I}$}{
					clear $\mathcal{K}$.\\
					\textbf{if} $G_{i,t}=0$: \\
					\quad Assign $\Omega[i] = L_t$.\\
					\quad \textbf{if} $ \frac{1}{M}\sum_{m=1}^{M}  \Omega[m] > L_\textrm{th}:$\\
					\qquad  Assign $\Omega[i] = L_{t-1}$.\\
					\qquad  \textbf{break} \\
					\quad Add $i$ to $\mathcal{K}$.\\
					Remove all elements of $\mathcal{K}$ from the index set, $\mathcal{I}$. \\
					\textbf{if} $\mathcal{I}$ is empty:\\
					\quad  \textbf{break} \\
			}}
			
		\end{small}
	\end{algorithm}
	
	We then investigate ways to solve (\ref{P3})  in order to achieve the group-wise adaptive design. Firstly, we denote the predicted $\boldsymbol{\delta} \big(\hat{\mathcal{D}}(\mathbf{s}_m,\mathbf{H}, B_m,\sigma^2)  > D_\textrm{th}\big)$ of the $m$-th sample choosing $L_t$ as the CSI codeword length as $G_{m,t}$. 
	The value of $\boldsymbol{\delta} \big(\hat{\mathcal{D}}(\mathbf{s}_m,\mathbf{H}, B_m,\sigma^2)  > D_\textrm{th}\big)$ is drawn from set $\left\{0,1\right\}$, which is determined by comparing the predicted $\hat{\mathcal{D}}\left(\mathbf{s}_m, \bH, B_m, \sigma^2\right)$ with $D_\textrm{th}$. To solve the problem, we propose a bottom-up water-filling algorithm. Specifically, we respectively search the $M$ images and assign the codeword length for each image by drawing the smallest but suitable $B_m$ from $\Lambda$. In this way, the higher codeword length will be prioritized to allocate to the images that are least likely to exceed the threshold, while images that cannot meet the distortion requirement for any compression level will be discarded. Moreover, when constraint (\ref{c2}) is unsatisfied, we terminate the algorithm and return the allocation results, $\Omega$. Then, the transmitter will send $\Omega$ represented by the indicator to the receiver to indicate the codeword length for each CSI matrix. The detailed procedure is summarized in Algorithm \ref{A5}.

	\begin{figure}[tbp]
		\begin{centering}
			\includegraphics[width=0.65\textwidth]{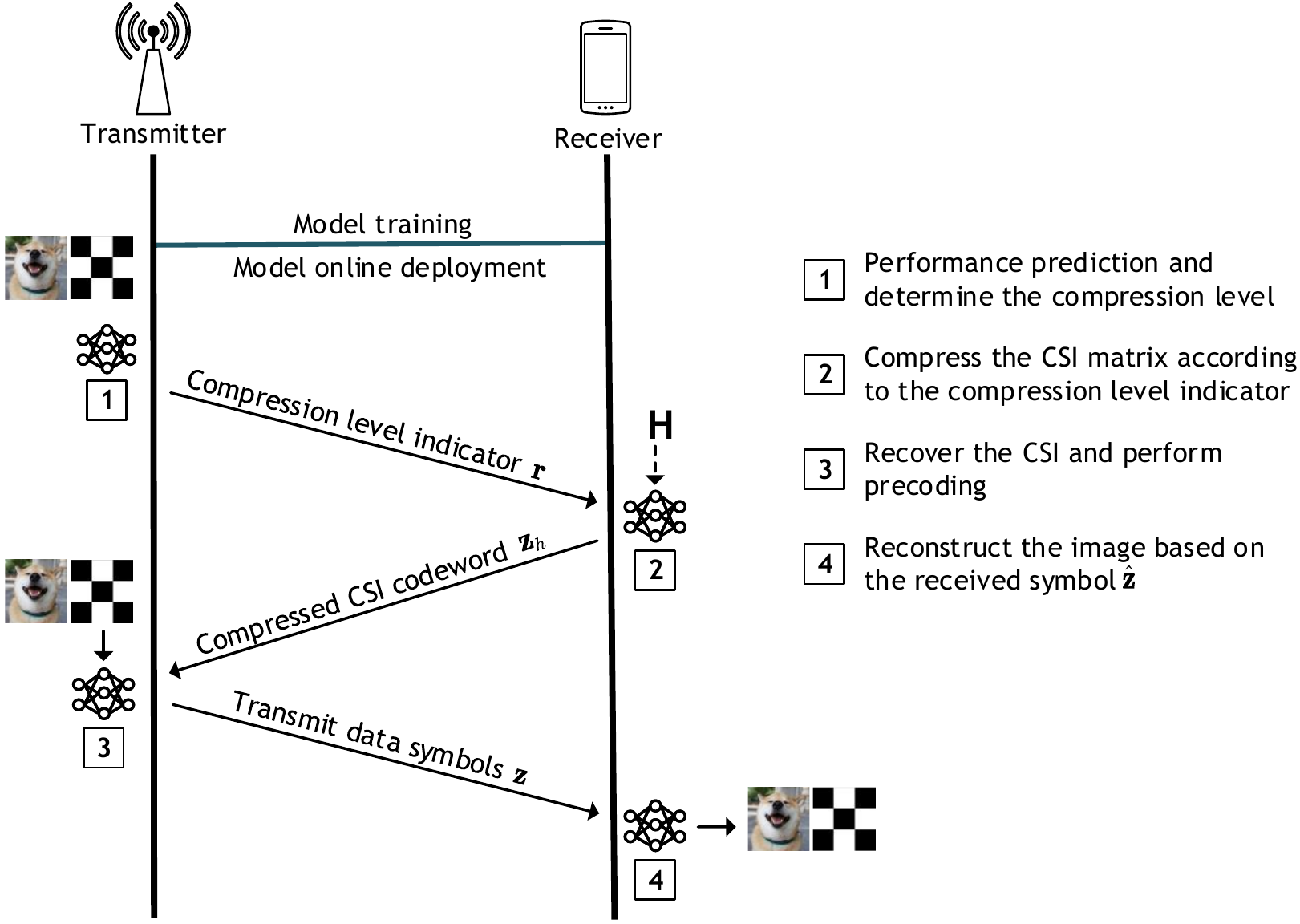}
			\par\end{centering}
		\caption{Communication process of SCAN.}
		\label{work_flow}
	\end{figure}
	\subsection{Communication Process of SCAN}
	We aim at adaptively adjusting the overhead of MIMO channel feedback for DeepSC-MIMO based on the predicted distortion. We train the proposed models, including DeepSC-MIMO, performance evaluator, and CSI model before deployment. The communication process of the semantic-aware adaptive channel feedback scheme is shown in Fig. \ref{work_flow}, and described as follows.
	\begin{itemize}
		\item[(i)]The transmitter determines the compression level based on the complexity of the images, and transmits the compression indicator, $\mathbf{r}$, to indicate the compression level, consisting of $\log_{2}L$ bits.
		\item[(ii)] The receiver compresses the estimated channel, $\mathbf{H}\in \mathbb{C}^{N_r \times N_t}$, according to the indicator and feeds back the CSI codeword to the transmitter.
		\item[(iii)] The transmitter recovers the CSI matrix with the CSI codeword to obtain the recovered $\hat{\mathbf{H}}$. It then transmits the precoded data symbols to the receiver by performing precoding with $\hat{\mathbf{H}}$. 
		\item[(iv)] The receiver reconstructs the image based on the received symbol vector $\hat{\mathbf{z}}$.
		
	\end{itemize}
	Furthermore, since we simultaneously determine the CSI codeword length for a group of images in the group-wise design, the transmitter will immediately send $M \log_{2}L$ bits to the receiver to indicate the compression levels of the $M$ images.

	\section{Simulation Results} \label{Simulation}
	\subsection{Simulation Setup}
	In the simulation, we consider a transmitter equipped with $N_t=16$ transmit antennas and a receiver equipped with $N_r=16$ receive antennas. The number of streams is $d=2$. We employ the popular narrowband millimeter wave (mmWave) clustered channel \cite{mmWave}. We implement the proposed DeepSC-MIMO and the deep learning-based channel feedback scheme with the deep learning platform “Pytorch”. The “Adam” optimizer is employed, with the batch size of $128$. Moreover, the initial learning rate is $0.0005$ and will be reduced with the increase of the number of epochs. We use the CIFAR10 dataset which consists of $50,000$ color images of size $32\times32\times3$ in the training dataset and $10,000$ images in the test dataset.
	
	As for channel feedback, we employ the CLNet \cite{CLNET}, which proposes a forged complex-valued input layer to process CSI data and utilizes spatial-attention to improve the performance. Given a codeword length, the encoder of CLNet compresses the CSI into a low-dimensional latent vector. Moreover, the optional codeword length is set as $\Lambda=[32,64,96,128,160,192]$, unless otherwise specified. To evaluate the performance of the DeepSC-MIMO, PSNR is chosen for distortion metric. 
	 It measures the ratio between the maximum possible power and the noise, which can be calculated by
	\begin{equation}
		\textrm{PSNR}=10 \log_{10}{\frac{\textrm{MAX}^{2}}{\textrm{MSE}}}(\textrm{dB}),
	\end{equation}
	where $\textrm{MSE}=d(\mathbf{s},\hat{\mathbf{s}})$ represents the mean square-error (MSE) between the source image, $\mathbf{s}$, and the reconstructed image, $\hat{\mathbf{s}}$. Moreover, $\textrm{MAX}$ is the maximum possible value of the pixels, e.g., $\textrm{MAX}$ equals $255$ for the images of RGB format. PSNR can be viewed as the inverse of the distortion. For comparison, we adopt the BPG source coding and the advanced low-density parity-check (LDPC) channel coding. The 16QAM and 4QAM are selected as the modulation schemes. Moreover, we also compare the DeepSC-MIMO with the classic DJSCC proposed in \cite{JSCC}. 
	
	\begin{comment}

 	For the simulation of $\mathbb{P}\left\{ \mathcal{E} \right\}$, we employ the Monte Carlo sampling, with the given a mini-batch of data $\{\mathbf{s}_i \}_{i=1}^{M}$, we have the following empirical estimation:
 	\begin{equation}\label{EM1}
 		\mathbb{P}\{\mathcal{E}\} \triangleq \frac{1}{M} \sum_{m=1}^{M} \left\{  \int_{\mathbf{z} \in \mathcal{E}\left(\mathbf{s}_m\right)} p_{\hat{Z} \mid Z}\left(\hat{\mathbf{z}} \mid \mathcal{F}_{\boldsymbol{\theta}}\left(\mathbf{s}_m\right)\right)d\mathbf{z} \right\}.
 	\end{equation}
 	The transmitter is only equipped with quantized CSI about instantaneous channel realizations and uses the quantized CSI for beamforming to improve the performance. Therefore, given the fixed channel realization, $\mathbf{H}$, and feedback length, $b$, the channel transition probability is only decided on the variance of AWGN. Therefore, we rewrite (\ref{EM1}) as 
 	\begin{equation}\label{EM2}
 		\mathbb{P}\{\mathcal{E}\} \triangleq \frac{1}{M} \sum_{m=1}^{M} \left\{  \int_{\mathbf{n}} p_{\mathbf{n}} \times \left(\mathcal{D}(\mathbf{s},\mathbf{H}, B_m,\mathbf{n})  > D_\textrm{th}\right) d\mathbf{n} \right\},
 	\end{equation}
 	Since we aim to investigate the effect of $B$, we sample the channel noise for each $\mathbf{s}_m$ $L$ times, then we have the further estimation:
 	\begin{equation}\label{EM3}
 		\mathbb{P}\{\mathcal{E}\} \triangleq \frac{1}{M} \sum_{m=1}^{M} \left\{   \frac{1}{L} \sum_{l=1}^{L} \left(\mathcal{D}(\mathbf{s},\mathbf{H}, B_m,,\mathbf{n}_l)  > D_\textrm{th}\right)   \right\},
    \end{equation}
\end{comment}

\begin{figure*}[!t]
	\centering
	\subfloat[]{\centering \scalebox{0.37}{\includegraphics{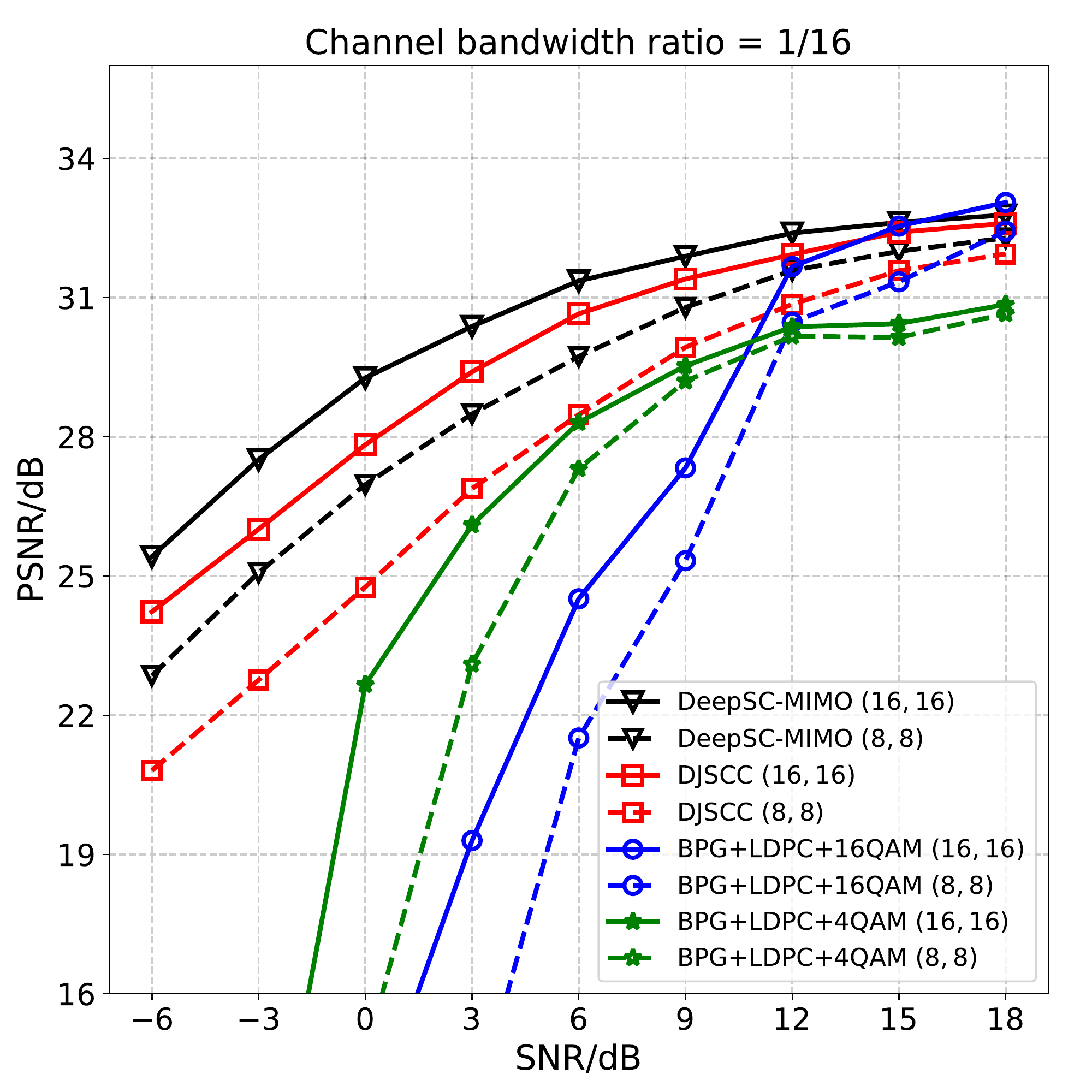}} }
	\subfloat[]{\centering \scalebox{0.37}{\includegraphics{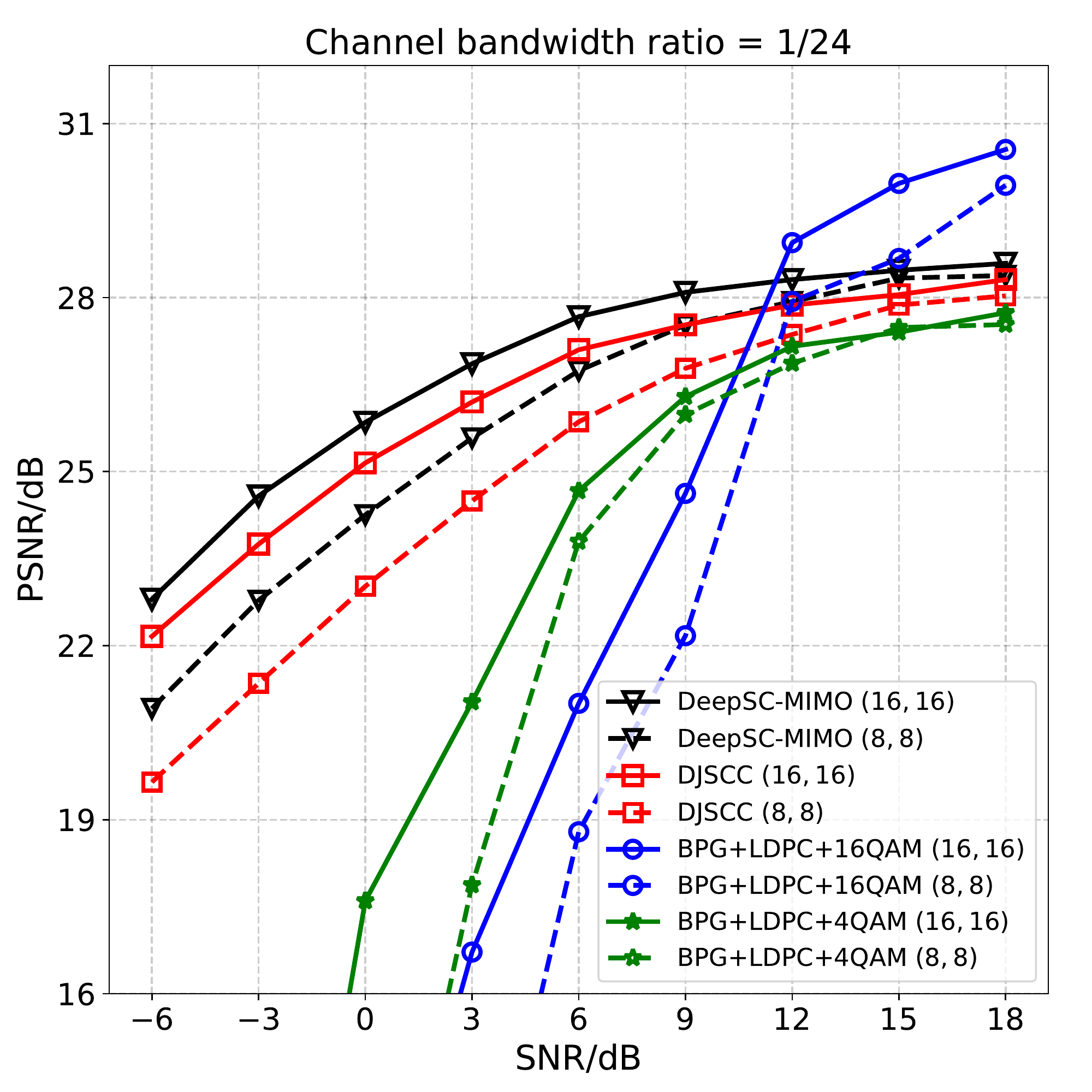}}}
	\caption{Performance versus SNR comparison of DeepSC-MIMO, DJSCC, and BPG+LDPC transmission methods under different numbers of antennas and channel bandwidth ratio. }
	\label{MIMO_Antenna}
\end{figure*}

	\subsection{Performance of DeepSC-MIMO}
	
	Fig. \ref{MIMO_Antenna} presents the performance of the investigated schemes equipped with different numbers of antennas versus the SNR and we assume that the transmitter obtains the perfect CSI. In addition, ``(16,16)" indicates $N_r=16$ and $N_t=16$, and the other notations can be understood in a similar way. We train the proposed DeepSC-MIMO model with SNR $=18$ dB and test it in SNR from $-6$ dB to $18$ dB. It is readily seen that the PSNR achieved by DeepSC-MIMO increases with SNR. The system equipped with more antennas generally outperforms the system with fewer antennas.  We also observe that DeepSC-MIMO can achieve relatively better performance than the standard separate coding scheme, especially in the low SNR regime. Besides, since we incorporate the CSI and channel noise into DeepSC-MIMO, it outperforms DJSCC in all SNR regimes. This is because the adaptive design is beneficial for exploiting the CSI information and allocating different power to different sub-channels. In addition, by comparing the results in Fig. \ref{MIMO_Antenna}(a) and Fig. \ref{MIMO_Antenna}(b), we find that the performance gain exists in different channel bandwidth ratios.
	
	\begin{figure}[!ht]
		\begin{centering}
			\includegraphics[width=0.45\textwidth]{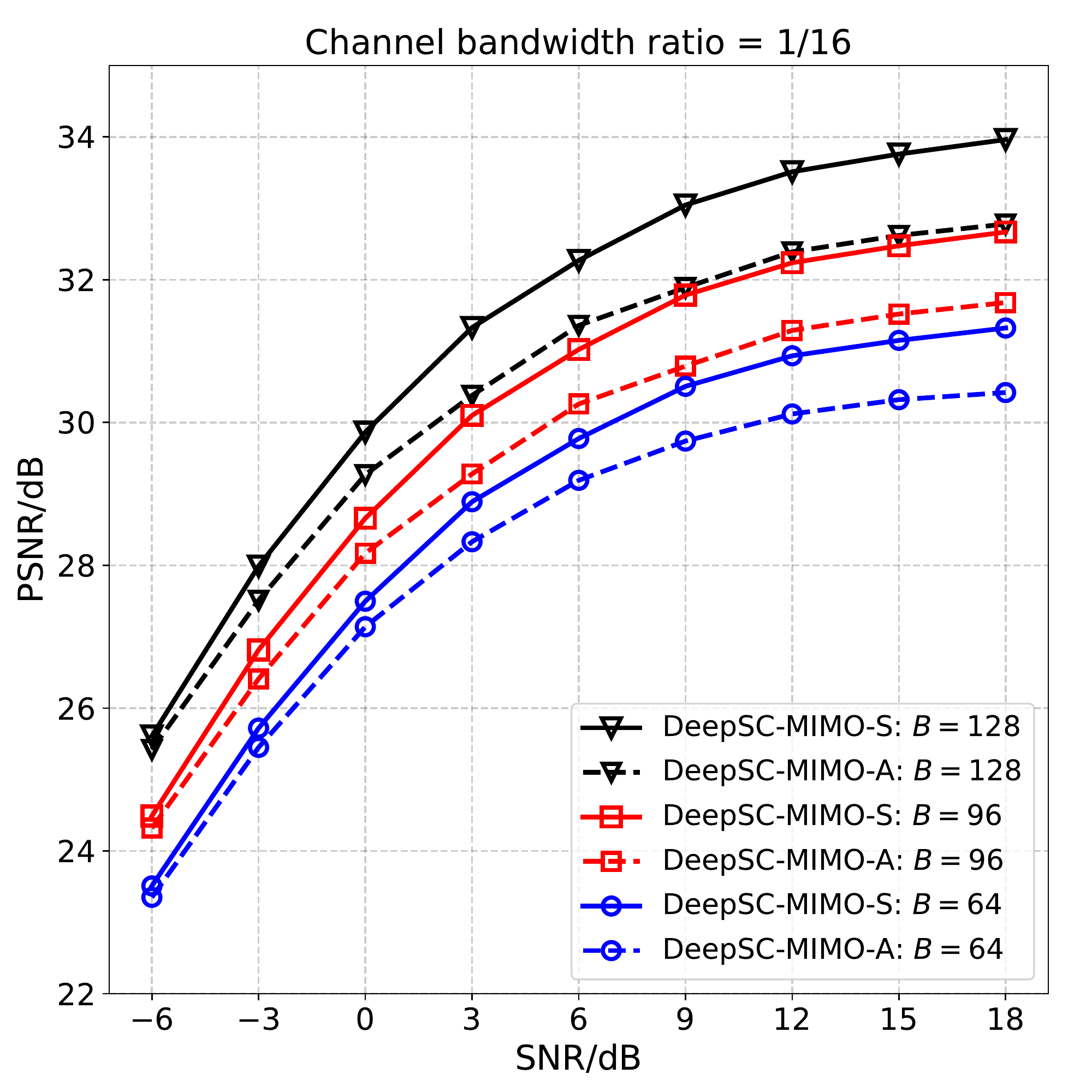}
			\par\end{centering}
		\caption{The performance of proposed DeepSC-MIMO at different SNR regimes and CSI compression levels.}
		\label{MIMO_AA}
	\end{figure}
	We also investigate the performance of adaptive design. In particular, we train DeepSC-MIMO with codeword length of $64$, $96$, and $128$, in the scenario where SNR is uniformly sampled from $[6,18]$ dB, and test the model in different SNR regimes. The results are given by DeepSC-MIMO-A in Fig. \ref{MIMO_AA}. In addition, the model trained in the fixed SNR is denoted as DeepSC-MIMO-S. It is demonstrated that there is a slight performance loss for DeepSC-MIMO-A compared with DeepSC-MIMO-S and the performance loss increases with SNR. Although there exists certain performance degradation, the adaptive design is still of significance in view of that it reduces the training time. Moreover, the performance gap between DeepSC-MIMO-A and DeepSC-MIMO-S increases with the codeword length. This is because the model is more sensitive to the disturbance in the case with high PSNR.

	\begin{figure*}[!t]
	\centering
	\subfloat[]{\centering \scalebox{0.2}{\includegraphics{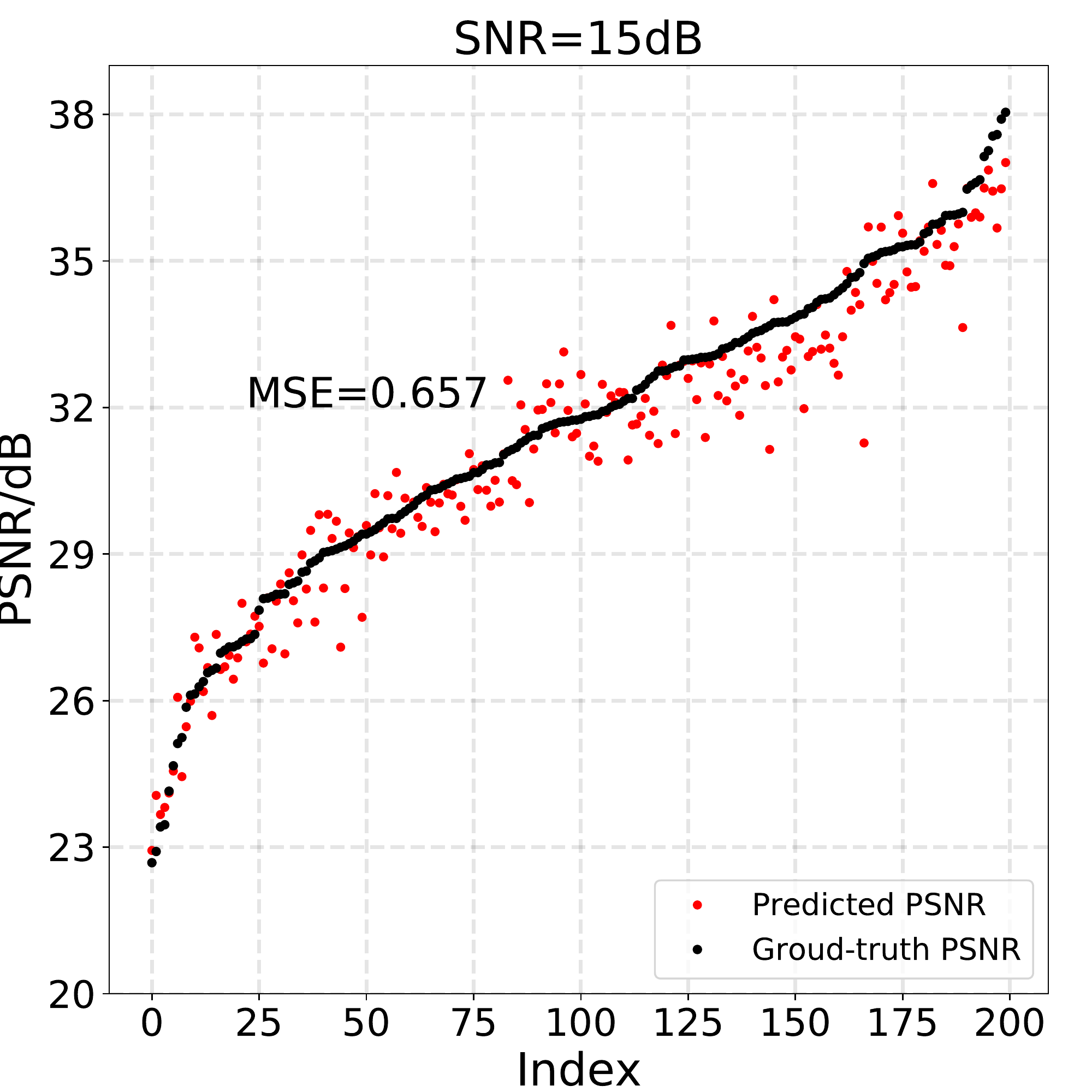}} }
	\subfloat[]{\centering \scalebox{0.2}{\includegraphics{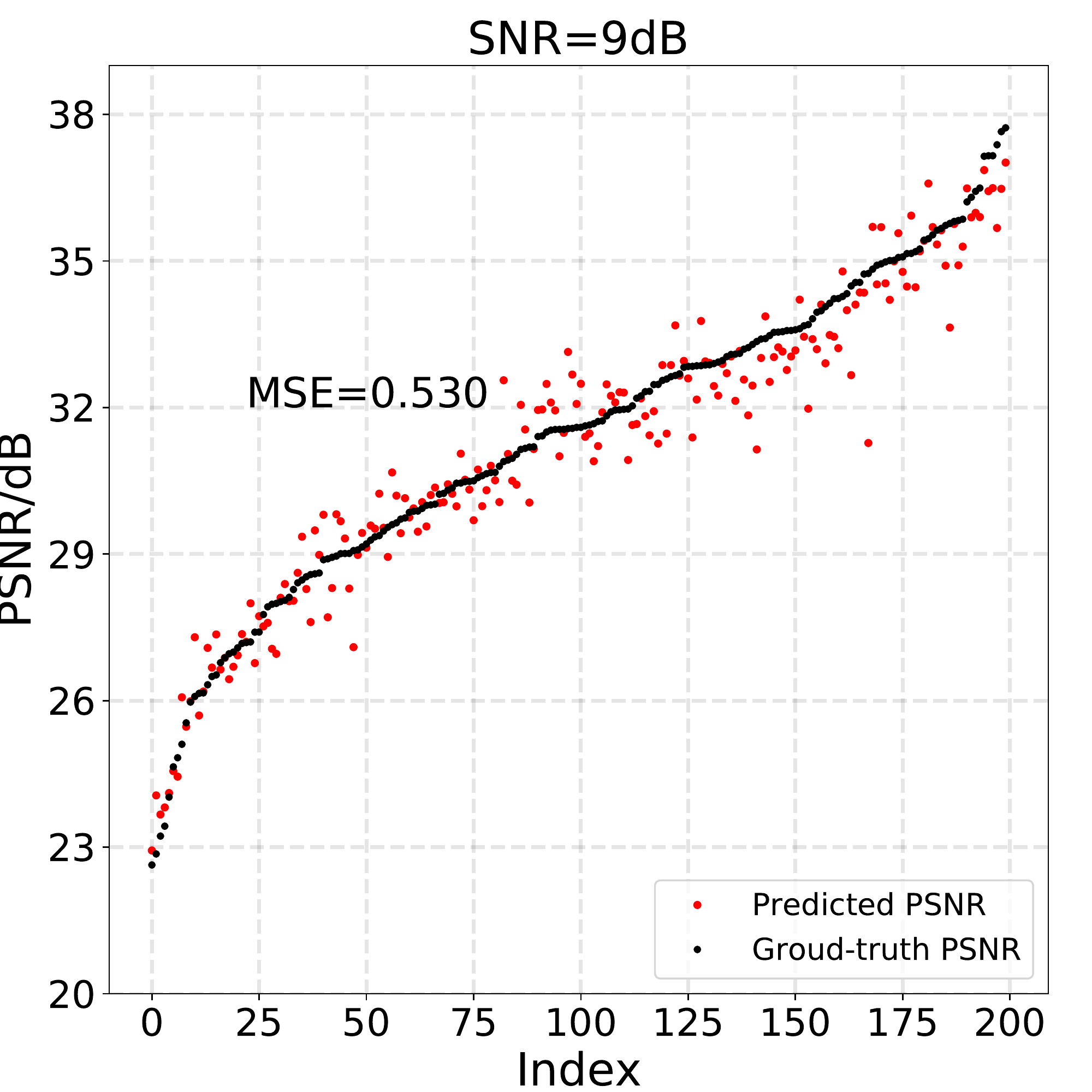}} }
	\subfloat[]{\centering \scalebox{0.2}{\includegraphics{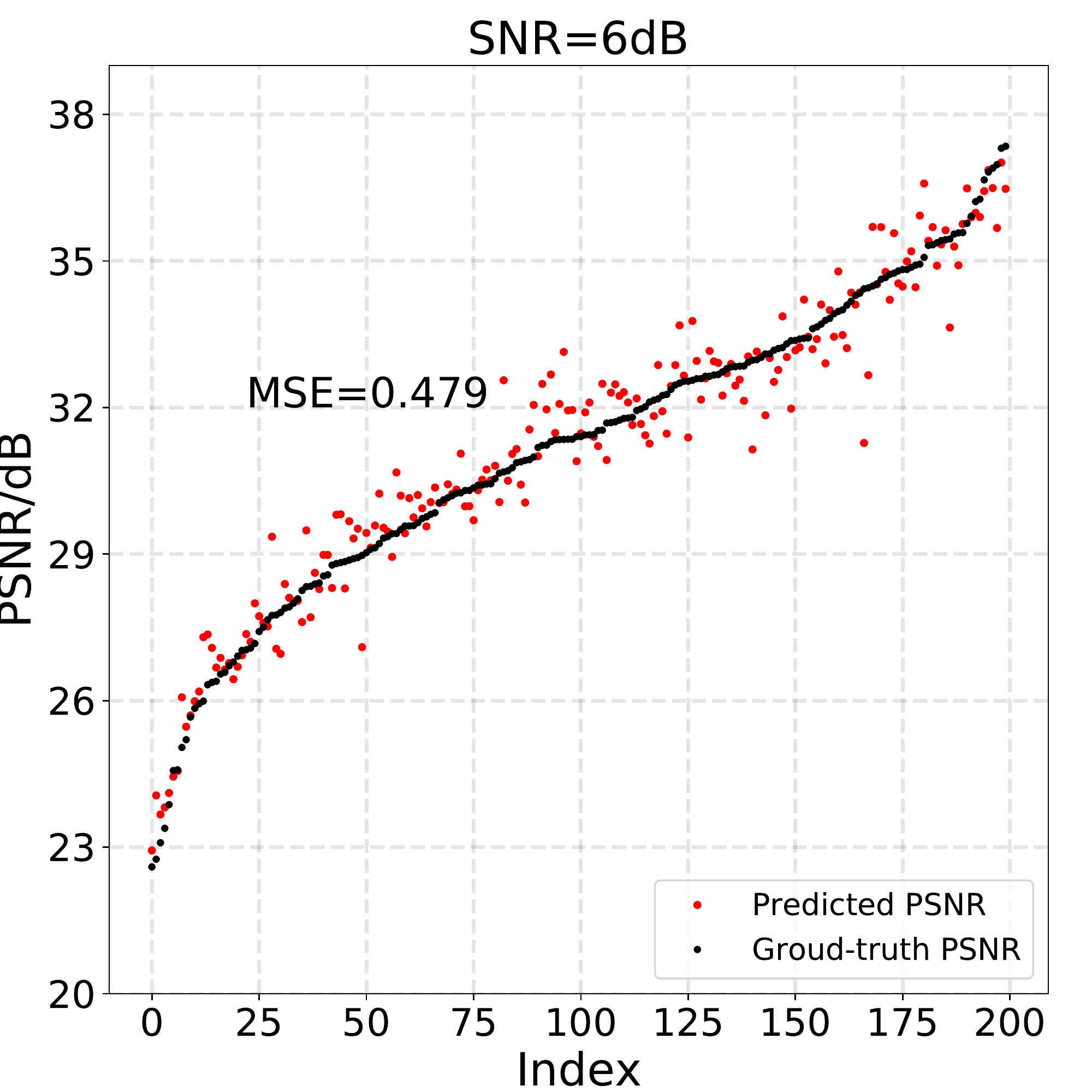}} }
	\subfloat[]{\centering \scalebox{0.2}{\includegraphics{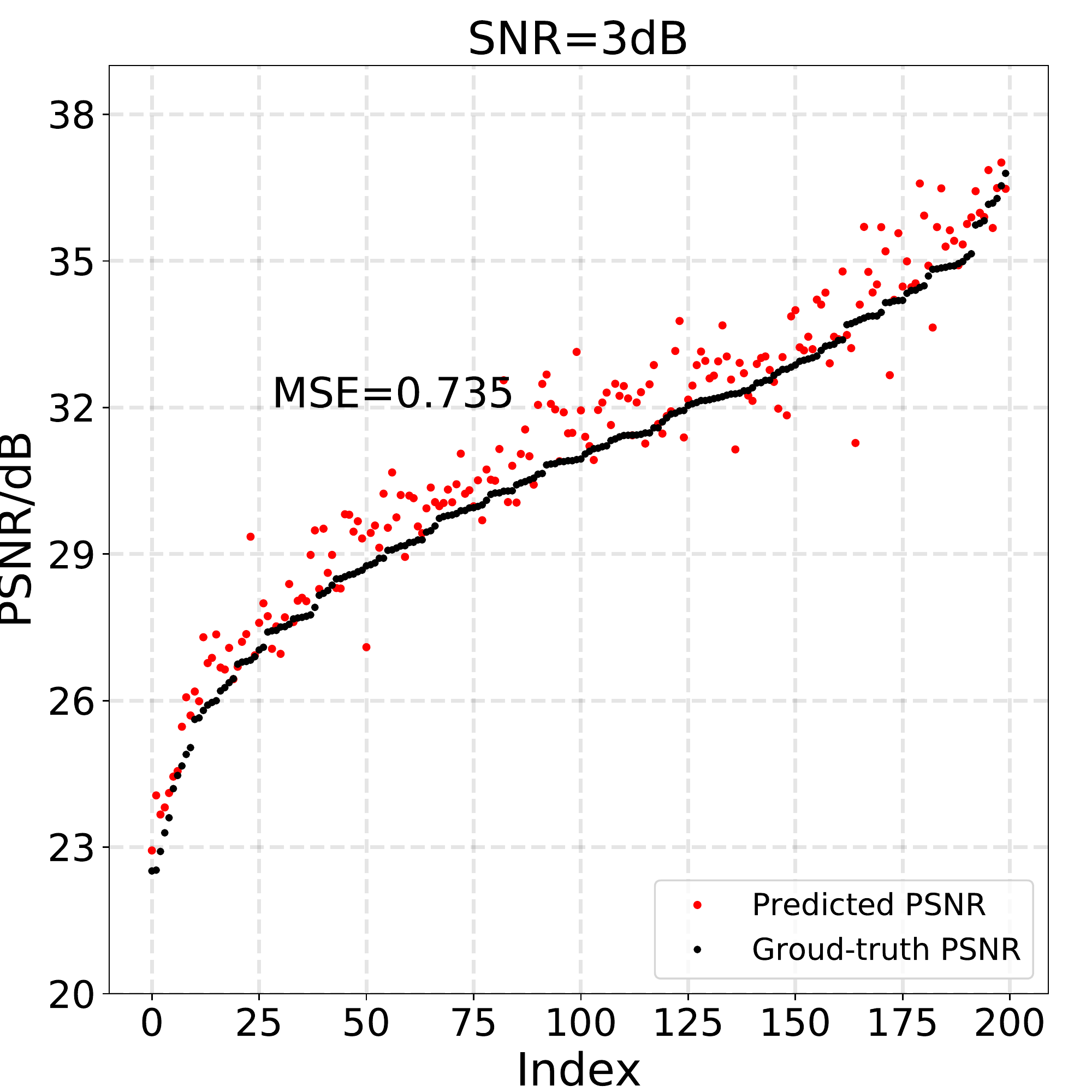}}}
	
	\subfloat[]{\centering \scalebox{0.2}{\includegraphics{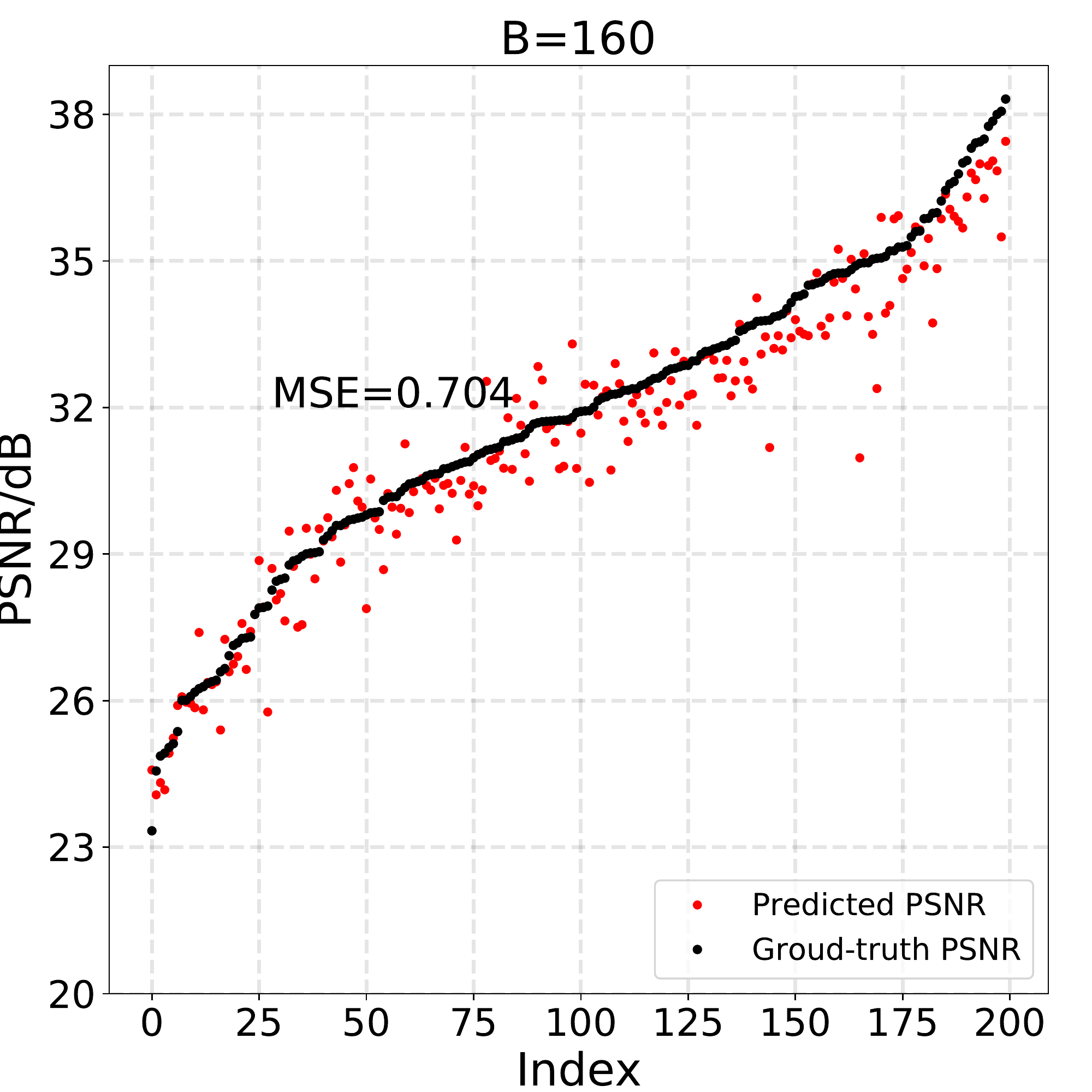}} }
	\subfloat[]{\centering \scalebox{0.2}{\includegraphics{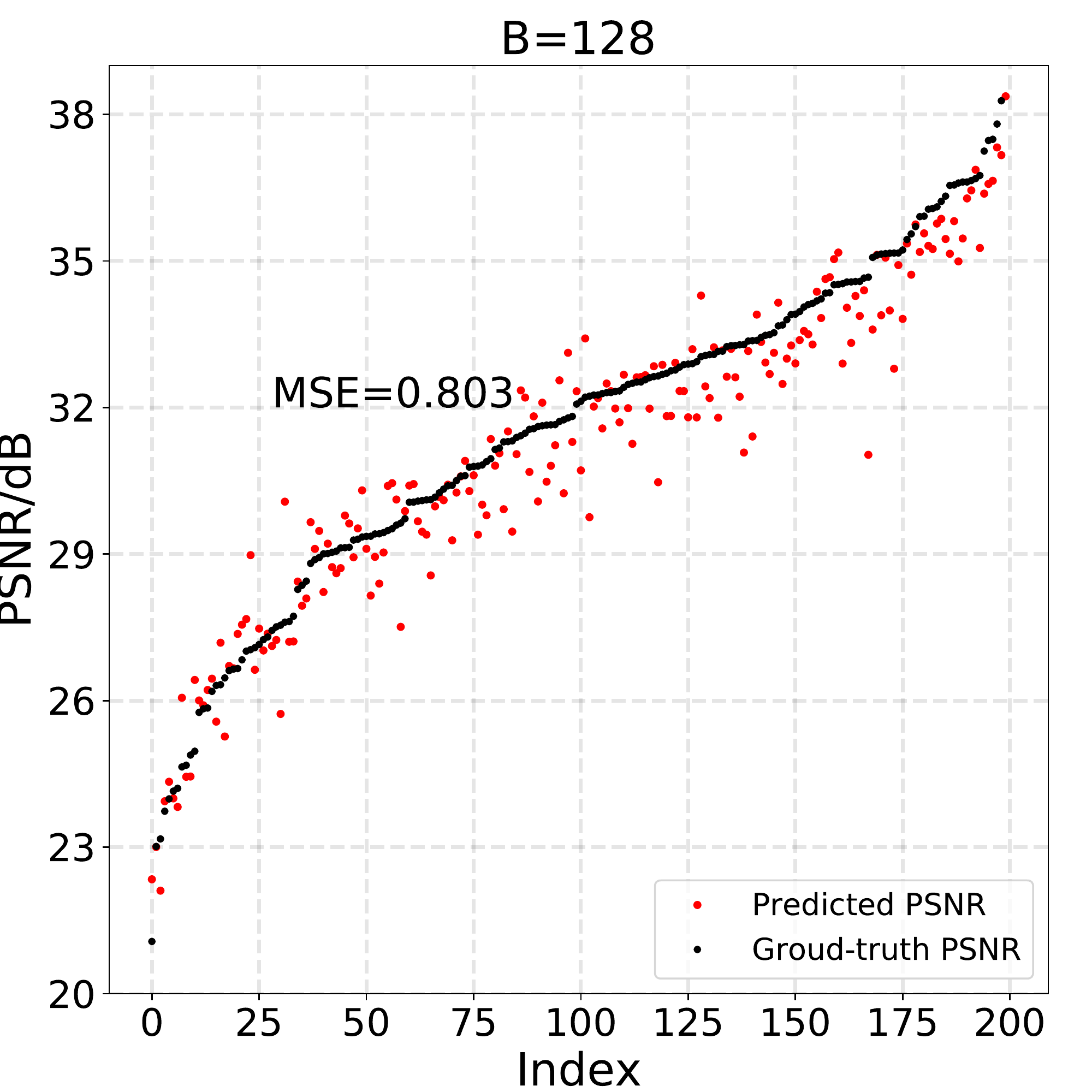}} }
	\subfloat[]{\centering \scalebox{0.2}{\includegraphics{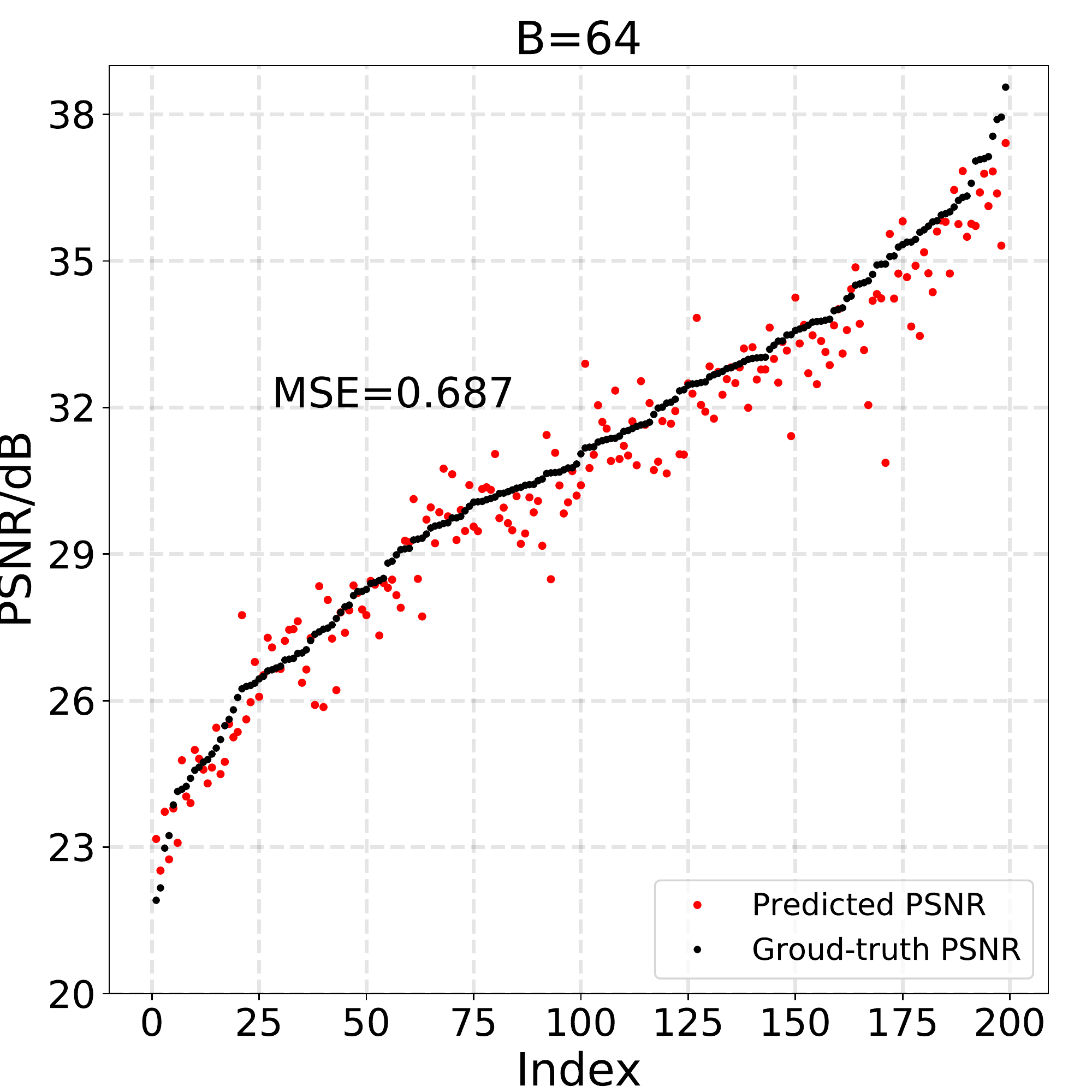}} }
	\subfloat[]{\centering \scalebox{0.2}{\includegraphics{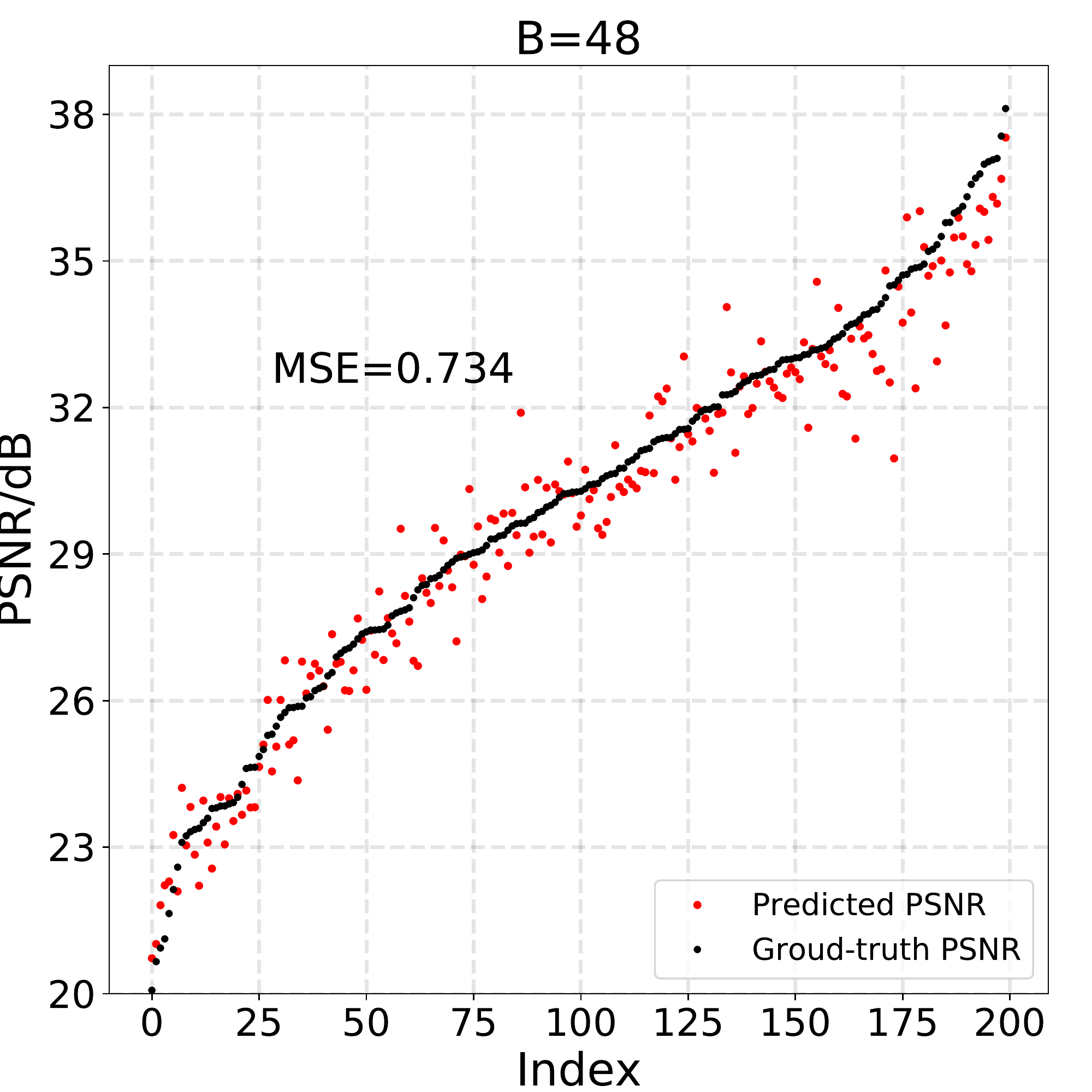}}}
	\caption{Prediction performance of the proposed performance evaluator. The first column shows the predicted PSNR value and the ground-truth PSNR value in different SNRs. The second column shows the predicted PSNR value and the ground-truth PSNR value in different CSI codeword length.}
	\label{PE_index}
\end{figure*}

	\subsection{Accuracy of Performance Evaluator}
	To evaluate the accuracy of the performance evaluator, we randomly select $200$ images from the test dataset and calculate the predicted PSNR values and the ground-truth PSNR values. The images are sorted from small to large according to the ground-truth PSNR values and the results are shown in Fig. \ref{PE_index}. From the figure, we observe that the performance evaluator can accurately predict PSNR values  with rather low prediction error. Besides, we compute the average prediction error on $200$ images as denoted by MSE marked in the figure. As we can see, the prediction error in the low SNR regime is larger than that in the high SNR regime. Moreover, the predicted PSNR of the performance evaluator is approaching the ground-truth values for the images with either high or low reconstruction quality. That is, the performance evaluator has successfully learnt to cope with the training bias of the model. According to the second column of Fig. \ref{PE_index}, the prediction error decreases with CSI codeword length. Most importantly, the reconstruction quality of these images are shown to be quite different from each other. Although the performance is close to average for the majority of images (about $70\%$), there are still some images where the reconstruction quality is rather low, introducing significant unreliability. Therefore, it is necessary to investigate the way to improve the performance of these images adaptively in semantic communications.  
	\begin{figure*}[!t]
		\centering
		\subfloat[]{\centering \scalebox{0.36}{\includegraphics{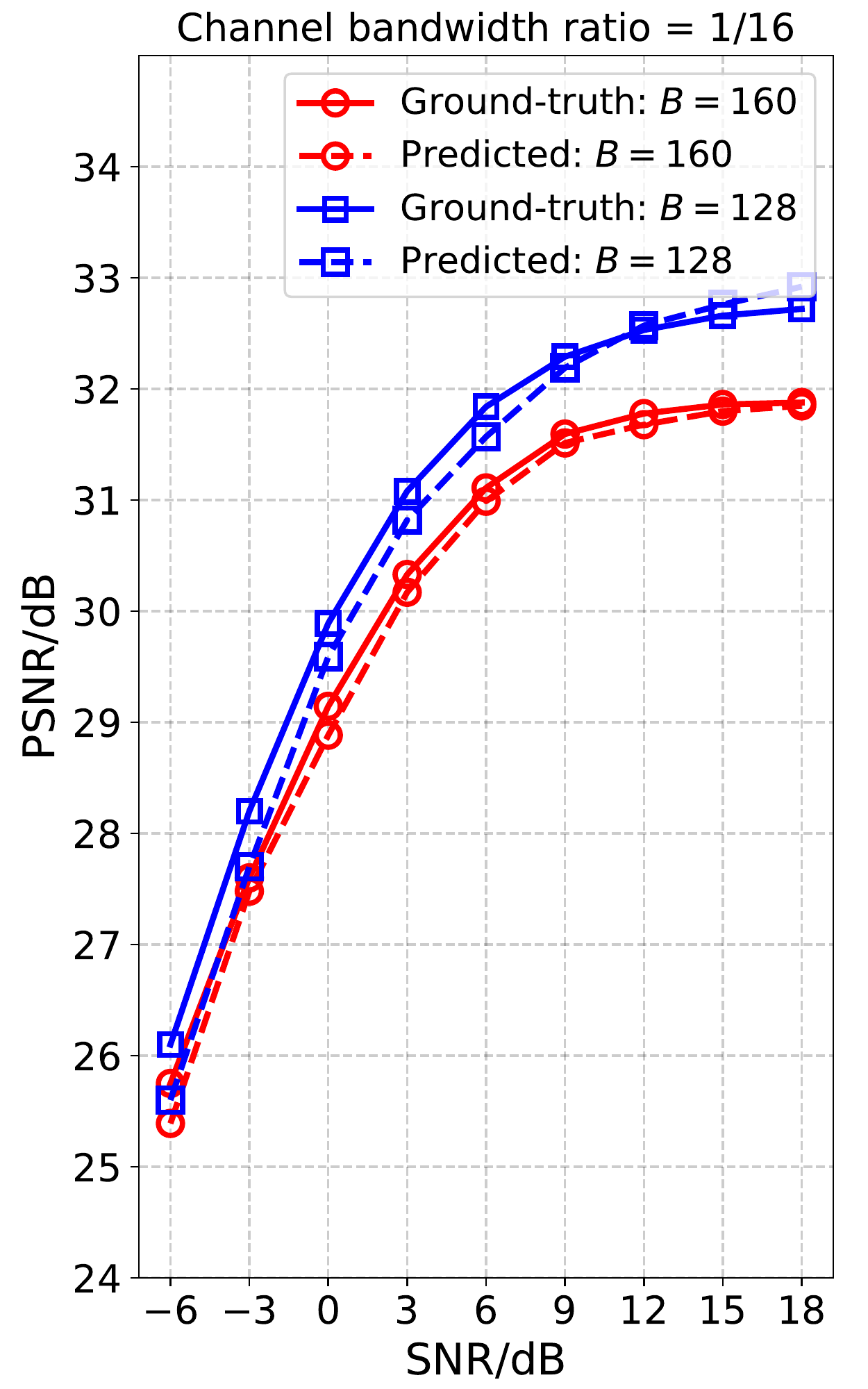}} }
		\subfloat[]{\centering \scalebox{0.36}{\includegraphics{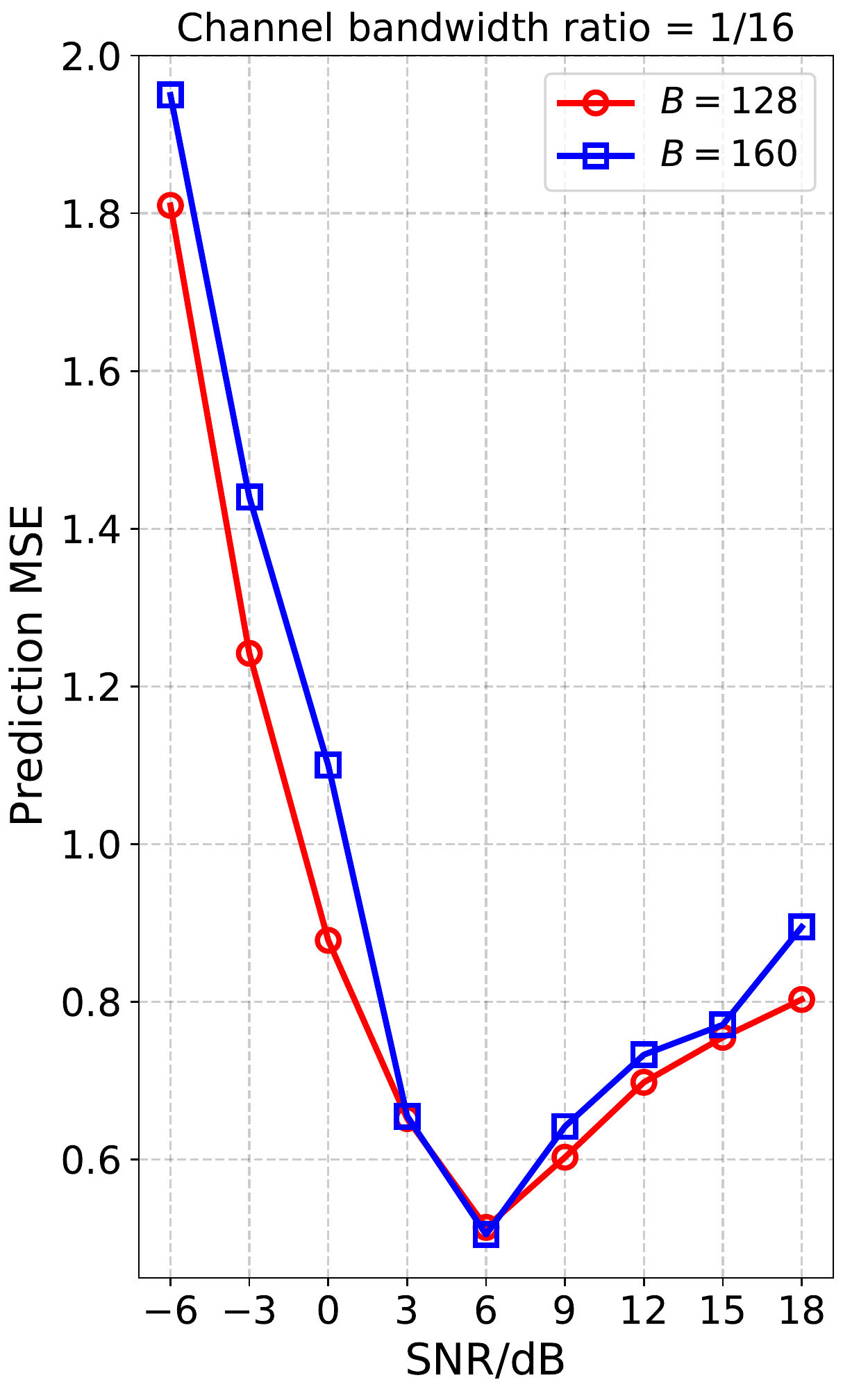}} }
		\caption{(a) The average ground-truth PSNR values and predicted PSNR values. (b) Prediction MSE of performance evaluator versus SNR.}
		\label{PE_SNR}
	\end{figure*}

	In Fig. \ref{PE_SNR}, we perform experiments on different CSI codeword length values. Specifically, in Fig. \ref{PE_SNR}(a), we compare the average ground-truth PSNR values and average predicted PSNR values achieved by DeepSC-MIMO and performance evaluator, respectively. We can observe that the predicted PSNR values are close to the ground-truth PSNR values. In Fig. \ref{PE_SNR}(b), we calculate the prediction MSE on different CSI codeword length values versus SNR. In general, we can see that the prediction MSE decreases with SNR, but the prediction MSE increases with SNR when SNR is lager than $6$ dB. This is because we sample the SNR from $-6$ dB to $18$ dB, where the average SNR is $6$ dB, Besides, when SNR is lager than $6$ dB, the ground-truth PSNR value will not significantly change with SNR. In this case, the performance evaluator tends to output similar results when SNR is lager than $6$ dB, confusing the training procedure. Moreover, at relatively higher SNR regime, PSNR value can be predicted accurately by the performance evaluator. In comparison,  the PSNR value is hard to predict at low SNR regime since the received features will be significantly disturbed by the noise, which makes the achieved PSNR vary over a large range.

		\begin{table}
		\centering  
		\caption{Comparison of model parameters.}    
		\label{Overhead} 
		\begin{tabular}{|c|c|c|c|c|}
			\hline
			Schemes & Prediction MSE & Model parameters     \\ \hline
			Basic model  & $0.524$ & $0.311$M    \\ \hline
			Small model       & $0.602$ & $0.142$M      \\ \hline
			Tiny model  & $0.704$ & $0.058$M     \\ \hline
			DeepSC-MIMO  & —— & $0.488$M      \\ \hline
		\end{tabular}
	\end{table}
	
	Table \ref{Overhead} compares the numbers of model parameters of different schemes. It is observed that the size of performance evaluator is much smaller than that of the DeepSC-MIMO, which shows the effectiveness of knowledge distillation. Besides, we also observe that the larger model is able to provide more accurate prediction results, hence the performance evaluator with proper parameters can be selected for practical uses.

	\begin{figure*}[!t]
		\centering
		\subfloat[]{\centering \scalebox{0.37}{\includegraphics{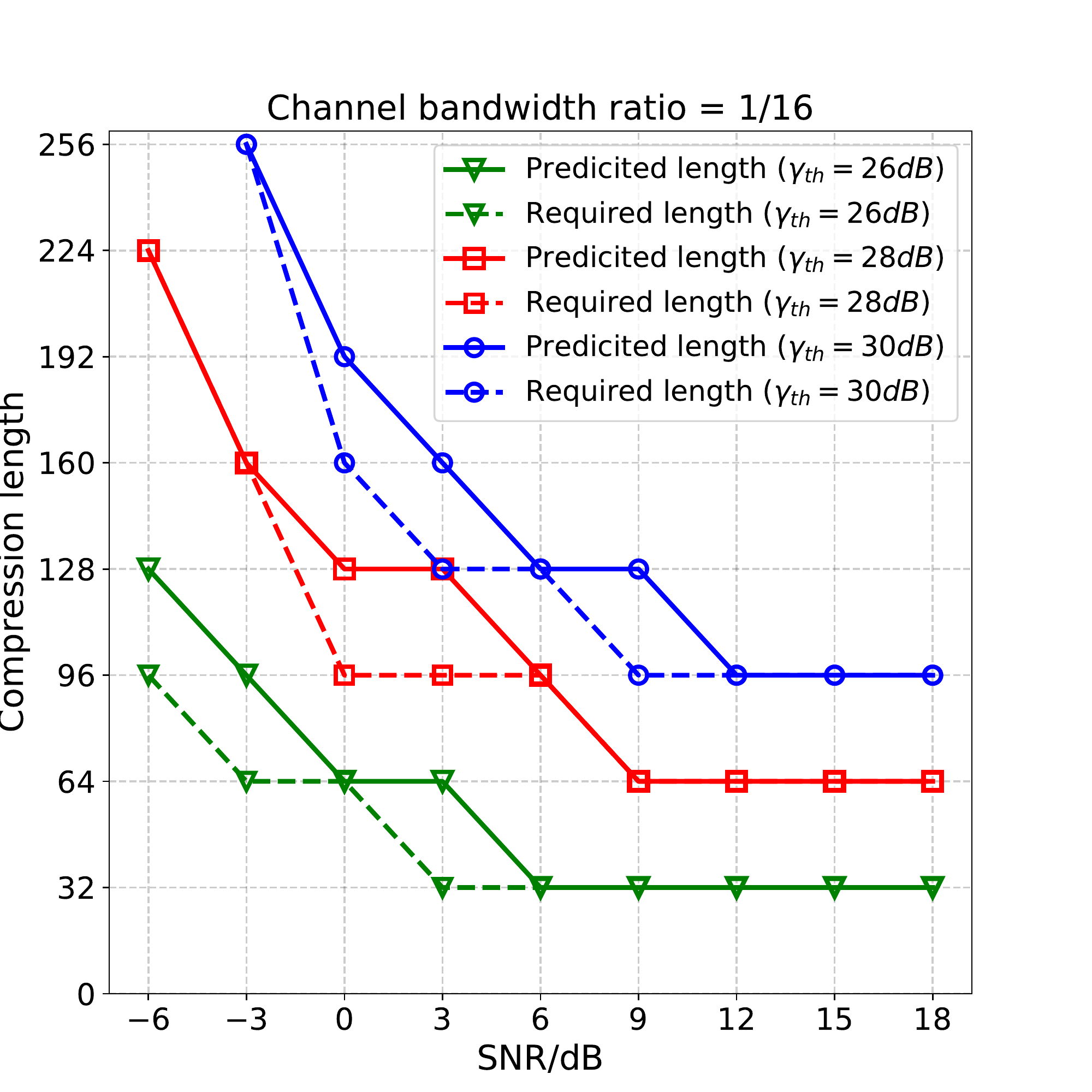}} }
		\subfloat[]{\centering \scalebox{0.37}{\includegraphics{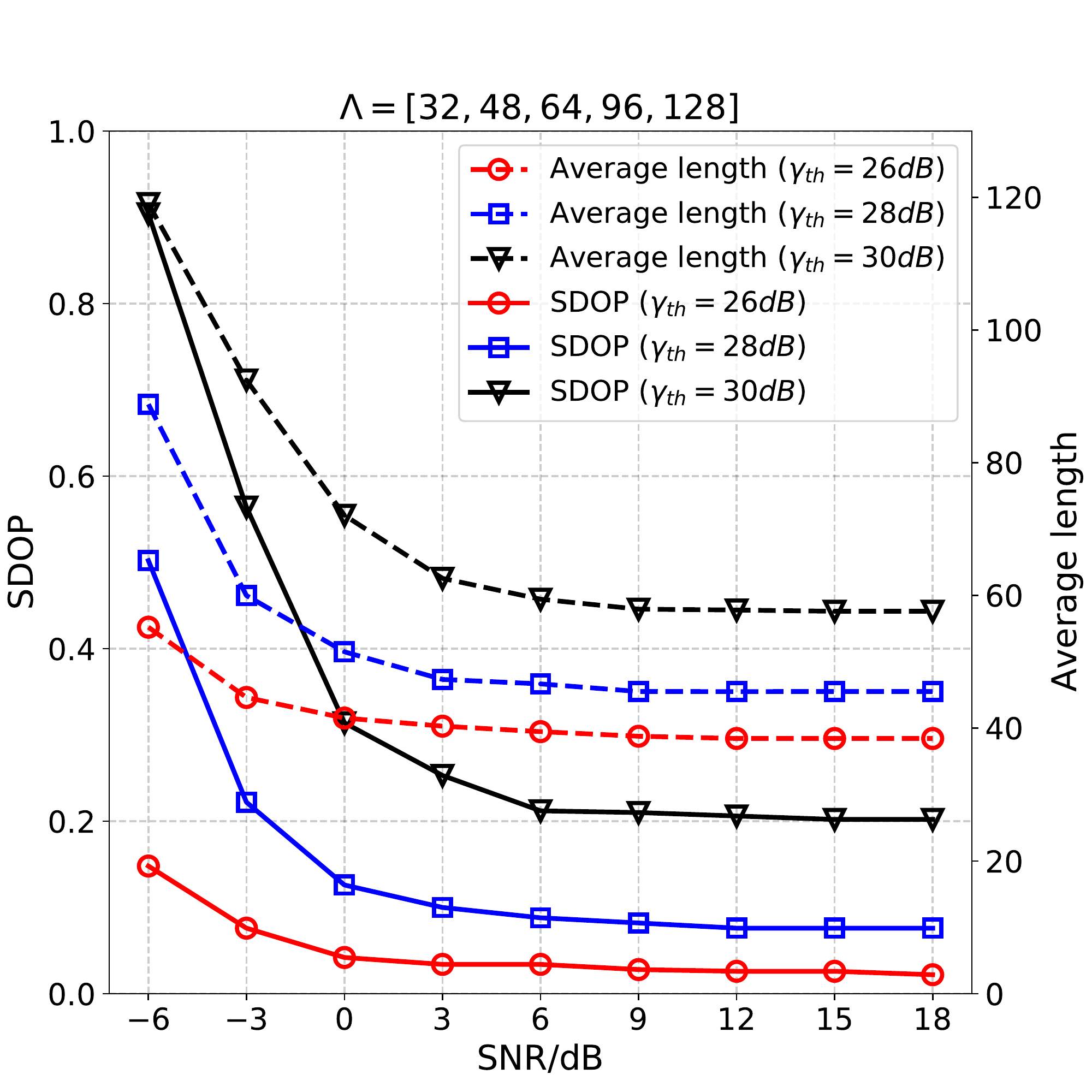}} }
		\caption{Performance of instance-wise SCAN. (a) The achieved codeword length when targeting for different PSNR thresholds. (b) The achieved SDOP and average required length when targeting for different PSNR thresholds.}
		\label{PE_pe}
	\end{figure*}

	\begin{figure*}[!t]
		\centering
		\subfloat[]{\centering \scalebox{0.37}{\includegraphics{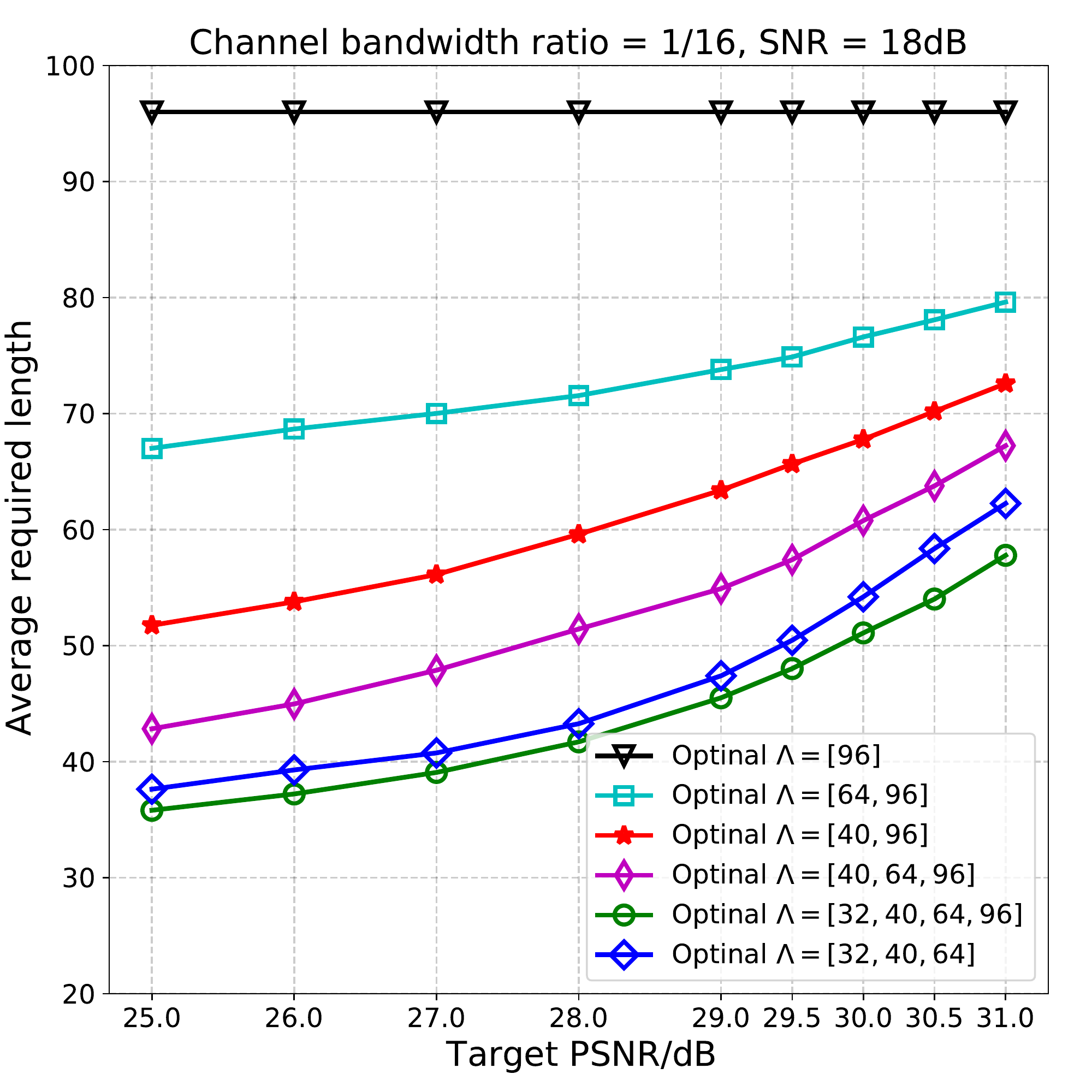}} }
		\subfloat[]{\centering \scalebox{0.37}{\includegraphics{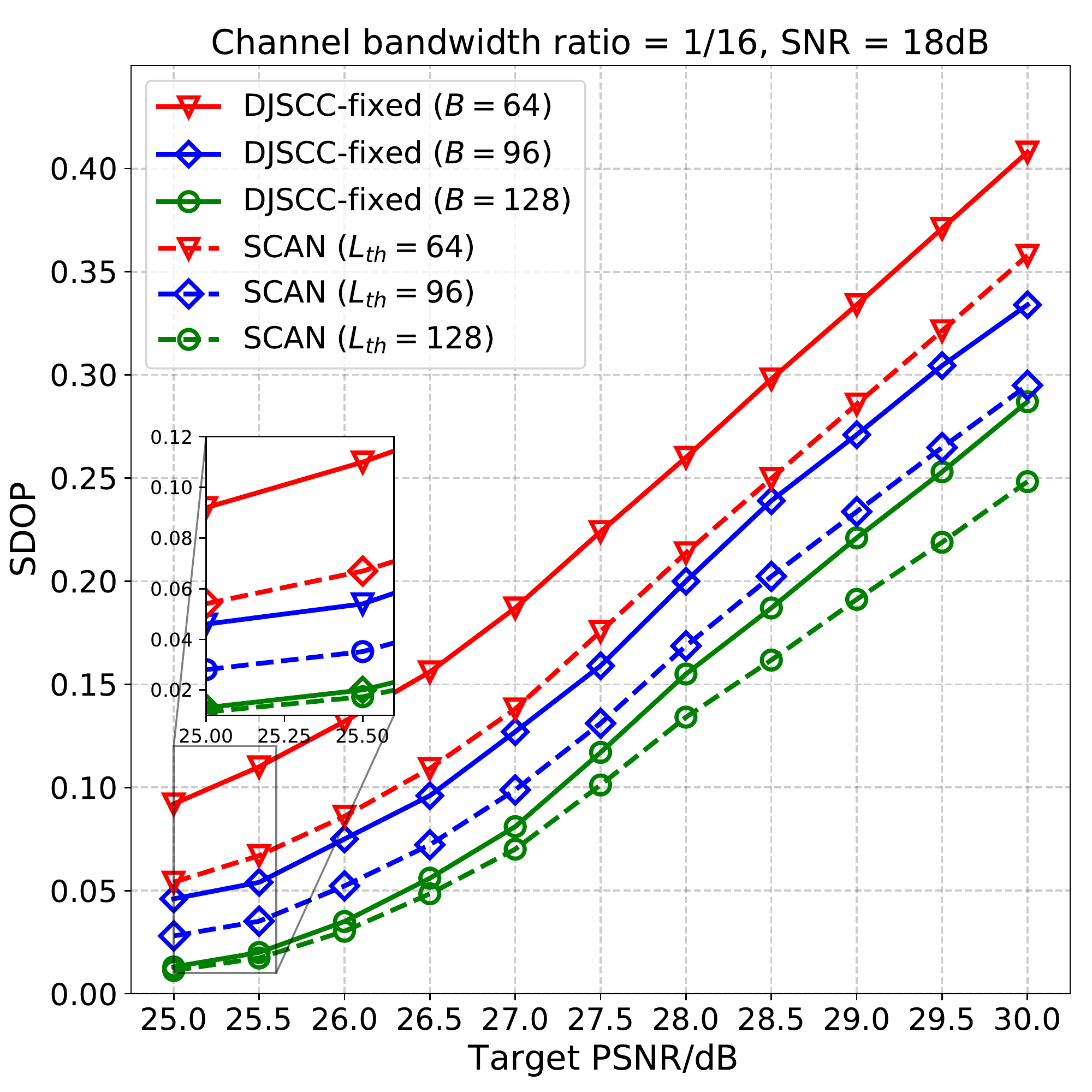}} }
		\caption{Performance comparison of group-wise SCAN. (a) Average codeword length comparison of SCAN with different optional length sets; (b) SDOP comparison between DJSCC and SCAN when using the same codeword length versus SNR.}
		\label{PE_TL}
	\end{figure*}
	\subsection{Results of SCAN}
	In this subsection, we aim to evaluate the performance of the proposed adaptive channel feedback scheme with respect to SDOP. The purpose is to adaptively adjust the channel feedback overhead according to the predicted performance. Particularly, we compare the achieved SDOP versus different target PSNR values, channel SNRs, and codeword length values. Moreover, the overhead of channel feedback is controlled by adjusting the compressed CSI codeword length, which is selected from the optional length set, $\Lambda$. 

	In Fig. \ref{PE_pe}, we present the performance of the instance-wise adaptive design. In particular, we randomly choose an image from the test dataset and consider $8$ codeword length values, i.e., $B \in \Lambda=[32,64,96,128,160,192,224,256]$. Then, we try to find the required minimal codeword length according to the ground-truth PSNR value. For comparison, we employ the performance evaluator to predict the PSNR value of this image and determine the predicted codeword length when targeting for different PSNR values, including $\bm{\gamma}_\textrm{th}=26$ dB, $\bm{\gamma}_\textrm{th}=28$ dB, and $\bm{\gamma}_\textrm{th}=30$ dB. The results are given in Fig. \ref{PE_pe}(a). From the figure, the performance evaluator can predict the optimal codeword length accurately, which shows that the proposed SCAN can determine the optimal CSI codeword length effectively. Moreover, it is also worth noting that the target can be unachievable for any CSI codeword length, e.g., $\bm{\gamma}_\textrm{th}=30$ dB at $\textrm{SNR}=-6$ dB. In Fig. \ref{PE_pe}(b), we calculate the overall SDOP when employing instance-wise scheme on all $10,000$ images. Since a higher $\bm{\gamma}_\textrm{th}$ is harder to achieve, the SDOP decreases significantly with $\bm{\gamma}_\textrm{th}$. Moreover, the average required CSI codeword length also decreases with SNR, which demonstrates that SCAN is a flexible scheme that enables variable rate feedback.
	
	In Fig. \ref{PE_TL}(a), we compare the average required codeword length when achieving the same SDOP. Specifically, all the schemes are required to achieve the same SDOP as employing $B=96$ for all the test images. From Fig. \ref{PE_TL}(a), it is readily seen that our proposed SCAN can significantly reduce the feedback overhead while keeping the same SDOP. Furthermore, it is shown that the feedback overhead can be reduced more with a larger codeword length set.
	
	To show the relative performance gain induced by the adaptive design, we compare SCAN and the DJSCC with fixed codeword length overhead scheme. That is, we employ the group-wise adaptive design and set the average length constraint, $B_\textrm{th}$, as $64$, $96$, and $128$, respectively. Then, we test the SDOP of DJSCC with fixed length by compressing the CSI for all images into the same length. From Fig. \ref{PE_TL}(b), it is observed that SCAN can achieve a significantly lower SDOP with the same average codeword length, demonstrating the superiority and flexibility of SCAN for realizing a more reliable semantic communication system.

\section{Conclusion} \label{Conclusion}
	In this paper, we proposed a novel metric, SDOP, to capture the reliability  of a semantic communication system. Then, to improve the reliability of a semantic communication, we developed a framework of SCAN including instance-wise and group-wise schemes, which are able to adjust the CSI codeword length based on the PSNR value of the image. To realize SCAN, we first proposed a semantic communication system, DeepSC-MIMO, for MIMO scenarios. We then developed a performance evaluator based on knowledge distillation, which can accurately predict the reconstruction quality of each image. Simulation results showed that the proposed scheme can significantly improve the performance and reliability with much reduced feedback overhead. Our proposed SCAN is a general framework and the future work could generalize it to other modalities of data. Within the framework, more adaptive designs, such as adaptive coding and adaptive modulation, can be potentially realized for higher performance gain. 
	
	%Regarding the reliability requirement of semantic communication and the poor generalization ability of DNNs, it is of great importance to explore efficient ways to enhance the reliability. Fortunately, the performance of the deep learning-based semantic communication can be evaluated, posing the potential to achieve the goal. This work has been the first step some open problems still need further study. For example, 

\bibliographystyle{IEEEtran}
\bibliography{IEEEabrv,Reference}

\end{document}